\documentclass{article}
\usepackage{arxiv}

\usepackage{times}
\usepackage{soul}
\usepackage{url}
\usepackage[hidelinks]{hyperref}
\usepackage[utf8]{inputenc}
\usepackage[small]{caption}
\usepackage{booktabs}
\usepackage[title]{appendix}
\usepackage{enumerate}
\usepackage{float}
\usepackage{subcaption}
\usepackage{setspace}
\usepackage{cite}
\usepackage{romannum}
\usepackage{tabularx}
\usepackage{array}
\usepackage{mdframed}

\usepackage{amsmath,amsthm,amssymb,amsfonts,bm}
\allowdisplaybreaks
\usepackage{mathtools}
\usepackage{stackrel}                                   
\usepackage{nicefrac}                                   
\usepackage{accents}
\usepackage{scalerel}

\usepackage{algorithmic}
\usepackage[ruled,vlined,linesnumbered]{algorithm2e}    
\newcommand{\nonl}{\renewcommand{\nl}{\let\nl\oldnl}}   
\usepackage{listings}

\usepackage{dsfont}

\usepackage{graphicx}
\usepackage{longtable}
\usepackage{hhline}
\usepackage{makecell}
\usepackage{varwidth}                                   
\usepackage{wrapfig}                                    
\newcommand{\halfspace}{\kern 0.2em}
\usepackage[explicit,compact]{titlesec}                         

\let\emptyset\varnothing

\usepackage[bottom]{footmisc}
\usepackage{xcolor}

\newcommand{\figref}[1]{Figure~\ref{#1}}

\newcommand{\R}{\mathbb{R}}

\newcommand{\Q}{\mathcal{Q}}
\newcommand{\opponentV}{U}
\newcommand{\playerV}{V}

\newcommand{\y}{\mathbf{y}}
\newcommand{\RSet}{\mathcal{R}}

\newcommand{\ic}[1]{\Theta_{#1}^{(i_{#1})}(\x_{#1-1})}
\newcommand{\ict}[1]{\Theta_t^{(i_t)}({#1})}
\newcommand{\icj}[2]{\Theta_{#1}^{(#2)}(\x_{#1-1})}
\newcommand{\icjx}[3]{\Theta_{#1}^{(#2)}(#3)}
\newcommand{\nsp}[2]{\mathcal{Q}_{#1}^{(#2)}}
\newcommand{\widehatic}[1]{\Theta_{#1}^{(i_{#1})}(\widehatx_{#1-1})}

\newcommand{\nodeset}{\mathcal{V}}
\newcommand{\edgeset}{\mathcal{E}}
\newcommand{\neighbor}[1]{\mathcal{N}_{#1}}

\newcommand{\inductivehypothesis}{\textit{Inductive Hypothesis: }}
\newcommand{\induction}{\textit{Induction Step: }}

\newcommand{\widehatx}{\widehat{\x}}
\newcommand{\widehatX}{\widehat{\X}}
\newcommand{\deltaX}[1]{\delta_{\X, #1}}

    \DeclareMathOperator*{\argmax}{argmax}                                      
    \DeclareMathOperator*{\argmin}{argmin}                                      
                                       
    \newcommand{\X}{\mathcal{X}}
    \newcommand{\x}{\mathbf{x}}
    \newcommand{\Y}{\mathcal{Y}}
    \newcommand{\K}{\mathcal{K}}
    \newcommand{\p}{\mathbf{p}}
    \newcommand{\q}{\mathbf{q}}
    \newcommand{\A}{\mathcal{A}}
    \newcommand{\z}{\mathbf{z}}
    
    \newcommand{\norm}[1]{\left\Vert #1 \right\Vert}
    \newcommand{\twonorm}[1]{\left\Vert #1 \right\Vert_2}
    \newcommand{\abs}[1]{\left\vert #1 \right\vert}
    
    \newcommand{\distH}[1]{\mathrm{dist}_{\mathrm{H}}\left(#1\right)}

    \usepackage{thmtools}
    \usepackage{thm-restate}

    \newtheorem{lemma}{Lemma}
    \newtheorem{assumption}{Assumption}

    \newtheorem{remark}{Remark}
    \newtheorem{definition}{Definition}
    
    \newtheorem{problem}{Problem}
    
    \graphicspath{{./figures/}}             

\title{On the Adversarial Convex Body Chasing Problem}

\author{
    Yue Guan \textsuperscript{1 *}\quad 
    Longxu Pan \textsuperscript{1 *} \quad 
    Daigo Shishika \textsuperscript{2} \quad
    Panagiotis Tsiotras \textsuperscript{1}
    \vspace{+3pt}
    \\
    \textsuperscript{1}
    Georgia Institute of Technology \quad
    \textsuperscript{2}
    George Mason University \vspace{+3pt}\\
}

\begin{document}

\maketitle
\begingroup\renewcommand\thefootnote{*}
\footnotetext{Equal contribution.}
\endgroup

\pagenumbering{arabic}

\begin{abstract}
    In this work, we extend the convex bodies chasing problem (CBC) to an adversarial setting, where an agent (the Player) is tasked with chasing a sequence of convex bodies generated adversarially by another agent (the Opponent).
    The Player aims to minimize the total cost associated with its own movements, while the Opponent tries to maximize the same cost. 
    The set of feasible convex bodies is finite and known to both agents, which allows us to provide performance guarantees with max-min optimality.
    Under certain assumptions, we show the continuity of the optimal value function,
    and propose an algorithm to numerically approximate the optimal policies for both the Player and the Opponent within a guaranteed tolerance.
    Finally, the theoretical results are verified through numerical examples.
\end{abstract}

\section{Introduction}

The Convex Bodies Chasing (CBC) problem was proposed in~\cite{friedman1993convex} to study the interaction between convexity and metrical task systems. 
It was soon realized that many problems of practical interest could be viewed as variants of the CBC problem, including scheduling~\cite{graham1966bounds}, efficient covering~\cite{alon2003online}, safe machine-learned advice~\cite{lykouris2018competitive,wei2020optimal}, self-organizing lists~\cite{sleator1985amortized}, the k-server problem~\cite{manasse1990competitive, koutsoupias1995k}, and other online convex optimization problems~\cite{hazan2016introduction}.
In the CBC problem, an \textit{online} agent (the Player) receives a request sequence of $T$ convex sets $\Q_1, \ldots, \Q_T$ contained in a normed space $\X$ of dimension $d$. 
The Player starts at $\x_0$ and, at time step $t$, observes the set $\Q_t$ and then moves to a new point $\x_t \in \Q_t$, which induces a cost $\norm{\x_t -\x_{t-1}}$. 
The objective of the Player is to maintain a constant ratio, known as the competitive ratio, against the minimum cost possible in hindsight, i.e., knowing the sequence of sets in advance. 
The existence of a finite competitive ratio was first conjectured in~\cite{friedman1993convex}.
Partial results on restricted cases were established later, including:
chasing subspaces~\cite{antoniadis2016chasing} and
chasing nested bodies~\cite{bansa2018nested,argue2019nearly}.
The conjecture was first resolved in~\cite{bubeck2019competitively}, which provided an $2^{\mathcal{O}(d)}$ upper bound. 
A nearly optimal competitive ratio was later derived in~\cite{bubeck2020chasing} for nested convex bodies using the classical Steiner point~\cite{schneider1971steiner}.
The more recent work~\cite{sellke2020chasing} has achieved a competitive ratio $\mathcal{O}(\sqrt{d \log T})$ without restrictions on the convex bodies, despite the fact that the proposed algorithm chooses $\x_t$ without knowledge of the future convex sets $\Q_{t+1},\ldots, \Q_T$.

In the classic CBC problem, with no restriction on the mechanism that generates the convex sets, the Player needs to select a point that balances the future cost for all possible subsequent convex sets.  
Consequently, the competitive \textit{ratio} is considered as the performance metric for most of the previous algorithms in the literature.
However, this performance metric can be ineffective in case of a high dimensional space $\X$.
Moreover, in many real-world scenarios the convex bodies are selected (potentially adversarially) from a known set of convex sets (e.g., dynamic Blotto game~\cite{shishika2021dynamic}). 
With this additional information, one expects to obtain better performance guarantees than with the competitive ratio.

In this work, we consider the \textit{adversarial} convex bodies chasing (aCBC) problem, where a (finite) set of compact convex bodies is known prior to starting the game, but the sequence of selected bodies is unknown to the Player, and is generated from the given set by an adversary (the Opponent). 
The adversarial selection of the convex bodies is further constrained over a graph, which implies that the currently selected convex body has an impact on the convex bodies available at the next time step\footnote{One can remove the graph constraint by using a fully-connected graph.}. 
The Player's movement is also constrained within the (compact) reachable set constructed from its current state.
We formulate this competitive game as a zero-sum sequential game~\cite{owen2013game} where the Player aims to minimize its total cost, while the Opponent tries to maximize it.

The contribution of this work is threefold: 
(i) we provide a novel formulation of the adversarial CBC game;
(ii) we provide theoretical guarantees for the existence of a Lipschitz continuous max-min value function under mild assumptions;
and (iii) we propose a numerical algorithm that provides bounded $\varepsilon$-suboptimal performance with respect to the max-min solution.

The rest of the paper is organized as follows:
Section~\ref{sec:formulation} formally presents the formulation of the adversarial convex bodies chasing problem;
Section~\ref{sec:value-function} introduces the optimal value function and provides theoretical results regarding its continuity\footnote{The terms ``optimal value function'' and ``value function'' are used interchangeably in this paper.}; 
Section~\ref{sec:algorithm} proposes a numerical algorithm that discretizes the domain and approximates the optimal policies for the Player and the Opponent. 
In the same section
we further prove that the total cost from the obtained policy is within $\varepsilon$-suboptimality of the optimal min-max solution; 
Section~\ref{sec:application} demonstrates the effectiveness of the proposed algorithm through numerical examples.
Finally, Section~\ref{sec:conclusion} concludes this work.

\section{Problem Formulation} \label{sec:formulation}
\begin{figure}[!htp]
    \centering
    \includegraphics[width=0.9\linewidth]{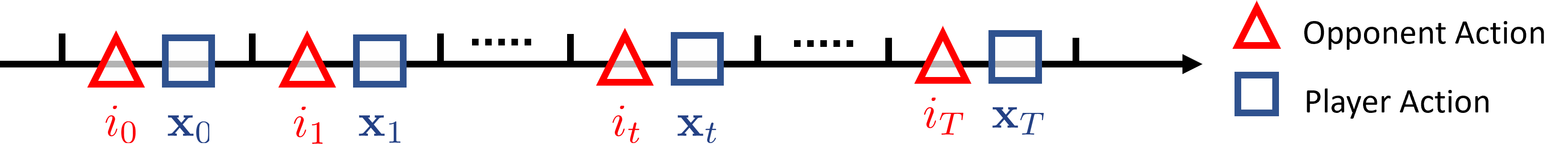}
    \caption{Timeline of the aCBC game. At each time step $t$, the Opponent first selects a state $i_t$ to assign a corresponding convex set $\nsp{t}{i_t}$ for the Player. 
    The Player then selects a point $\x_t$ within the assigned convex set.}
    \label{fig:timeline_demo}
\end{figure}

%
%
%
The adversarial convex bodies chasing (aCBC) \textit{game} is played sequentially between two agents: the Player and the Opponent.
The game evolves over a compact subset $\X$ of a normed Euclidean space $\mathbb{R}^d$.
At each time step $t$, a convex region $\Q_t \subseteq \X$ is selected by the Opponent, and then the Player chooses a point $\x_t \in \Q_t$, inducing a corresponding cost.
Once $\x_t$ is chosen, the Opponent selects the next convex set $\Q_{t+1}$ and the process continues until finite time horizon $T$.
The timeline of the game is presented in \figref{fig:timeline_demo}.
The Player tries to minimize its total cost $\sum_{t=0}^{T-1} c(\x_t, \x_{t+1})$ over a finite horizon $T$ for some non-negative cost function $c$, while the Opponent aims to maximize this cost. 

We use $\x_t$ to denote the state of the Player at time step $t$ and treat $\X$ as the state space of the Player.
Different from the classical CBC problem, we restrict the Player's selection of its next state to a neighborhood of its current state characterized by a \textit{reachability correspondence} (a set-valued map).
Specifically, we require that $\x_{t+1} \in \RSet(\x_t)$ for all $t=0,\ldots,T-1$, where $\RSet(\x_t)$ represents the \textit{reachable set} of the Player at the next time step from the current state $\x_t$.
We use $\RSet: \X \rightsquigarrow \X$ to denote the reachability correspondence $\RSet$ as a set-valued 
map\footnote{One can also use $\RSet: \X \to 2^{\X}$ to denote the reachability correspondence $\RSet$ as a single-valued map.}.



\begin{assumption}   \label{assmp:R set compact}
    For all $\x \in \X$, $\RSet(\x)$ is solid, compact, and convex.
\end{assumption}
At time~$t$, the Opponent's state is defined as the node $i_t \in \nodeset$ in a directed graph $\mathcal{G}=(\nodeset, \edgeset)$ that constrains the Opponent movements. 
Given the Opponent's current state $i_t$, the Opponent can move to any one of the neighboring nodes $i_{t+1}$, such that $(i_t, i_{t+1}) \in \edgeset$.
We denote the set of all neighbors of node $i_t$ as $\neighbor{i_t}$.
Before the game, we assign a finite collection of convex regions $\Q = \big\{\nsp{t}{i}\big\}_{t=0, i=1}^{T, |\nodeset|} \subset \mathcal{X}$ to the Opponent.
At each time step $t$,
the Opponent selects a convex region $\nsp{t}{i_t}$ from $\Q$ by selecting the feasible Opponent state $i_t \in \neighbor{i_{t-1}}$ from its previous state $i_{t-1}\in \nodeset$.
%
In other words, instead of having the freedom to choose an arbitrary convex subset of $\X$ as in the classical CBC problem, the Opponent in an aCBC game can only choose from a given finite set of convex bodies by selecting the next state (node of $\mathcal{G}$) to visit.
We make the following two assumptions on $\Q$ and the information structure of the game.
%
\begin{assumption}
    \label{assmp:convex body}
    For all $i \in \nodeset$ and $t=0,\ldots,T$, the set $\nsp{t}{i}$ is solid, compact, and convex.
\end{assumption}

\begin{assumption}
    \label{assmp:info}
    The collection $\Q$ and the graph $\mathcal{G}$ are common knowledge to both agents prior to the game.
\end{assumption}

Once a convex region $\nsp{t}{i_t}$ is selected at time $t$ by the Opponent, the Player needs to move to a feasible point $\x_t \in \RSet(\x_{t-1}) \cap \nsp{t}{i_t}$.
For ease of notation, we introduce the following intersection correspondence
\begin{equation}
    \label{eqn:intersection-correspondence}
    \ic{t} = \RSet(\x_{t-1}) \cap \nsp{t}{i_t}.
\end{equation}

To avoid degeneracy, we assume that any admissible sequence of convex bodies chosen by the Opponent is always feasible for the Player.
Consequently, the aCBC game is an optimization problem rather than a feasibility problem, similar to the classical CBC problem.

\begin{assumption}
    \label{assmp:feasibility}
For all $t = 1, \ldots, T$ and $i_t \in \nodeset$, the following holds:
    \begin{equation}
        \mathrm{int}\Big(\RSet(\x) \cap \nsp{t}{i_t}\Big) \neq \emptyset, \quad 
        \forall \; \x \in \nsp{t-1}{i_{t-1}}, \text{ where } i_t \in \neighbor{i_{t-1}}.
    \end{equation}
\end{assumption}

\begin{figure*}[!htp]
    \centering
    \includegraphics[width=0.85\textwidth]{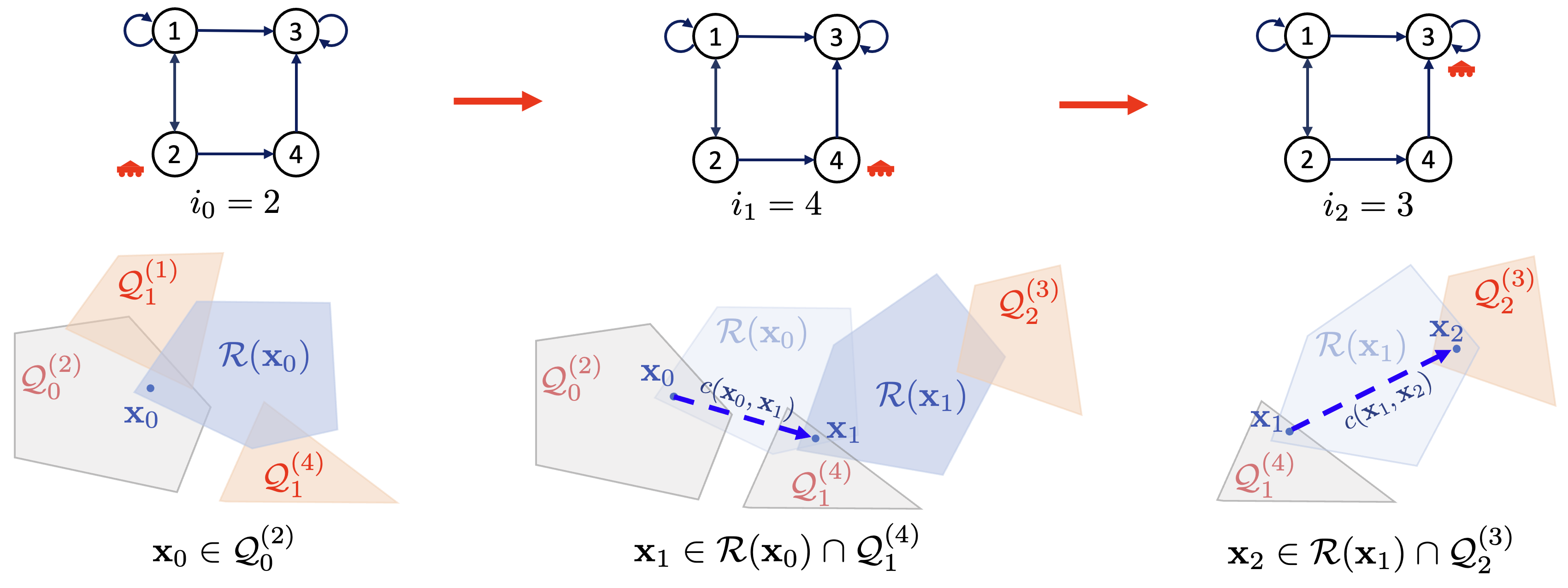}
    \caption{Example of an aCBC game with a 4-node graph and  a two-dimensional Player state $\X$.
    At time step $0$, the Opponent starts at node 2 and the Player selects a point $\x_0$ in $\nsp{0}{2}$. 
    The Opponent then moves to node $4$ and the Player selects the point $\x_1 \in \RSet(\x_0) \cap \nsp{1}{4}$, inducing a cost of $c(\x_0,\x_1)$.
    The Opponent then moves to node $3$ and the Player moves to $\x_2 \in \RSet(\x_1)\cap \nsp{2}{3}$ incurring a cost of $c(\x_1,\x_2)$, and so on.
    The game continues until reaching the finite time horizon $T$. 
    }
    \label{fig:aCBC_demo}
    \vspace{-10pt}
\end{figure*}

Assumption~\ref{assmp:feasibility} ensures that the optimization problem faced by the Player is strictly feasible for all time steps. 
With Assumptions~\ref{assmp:R set compact}-\ref{assmp:feasibility}, the major difference of the aCBC from the classic CBC formulation is that
(i) the selection of a new convex body considers the feasibility from the previous convex body, (ii) the feasible convex bodies $\Q_t^{(i)}$ are compact and solid, and (iii) the set of all convex bodies in the aCBC game is \textit{finite} and is \textit{common knowledge} to both agents.
An illustrative example of the proposed aCBC game is shown in ~\figref{fig:aCBC_demo}. 

The assumptions on the compactness of the convex bodies and the finiteness of its collection $\Q$ significantly reduce the Opponent's freedom of selecting convex regions.
This allows the construction of a min-max solution that is discussed in the next section. 
To ensure the existence of a min-max solution, however,
we need to further make the following assumptions on the continuity of the cost function $c$ and the reachability correspondence $\RSet$.

\begin{assumption}    \label{assmp:cost cont}
    The cost function $c: \X \times \X \to \mathbb{R}$ is 
    continuous.
    %
    %
\end{assumption}
Notice that the cost function here can be an arbitrary continuous function and thus is more general than the norm-cost used in the classic CBC formulation.
%
%
To distinguish the continuity of correspondences from the continuity of single-valued maps, we need to introduce the concepts of lower and upper semi-continuity.

\begin{definition}[upper semi-continuity \cite{freeman2008robust}]
    A set-valued map $F: \X \rightsquigarrow \Y$ is upper semi-continuous (usc) at $\x \in \X$ if for every open set $U \subseteq \Y$ such that $F(\x) \subseteq U$, there exists a neighborhood $V$ of $\x$ such that $F(V) \subseteq U$. $F$ is usc on $\X$ if it is usc at every point in $\X$.
    \label{def:set-valued map usc}
\end{definition}

\begin{definition}[lower semi-continuity \cite{freeman2008robust}]
A set-valued map $F: \X \rightsquigarrow \Y$ is lower semi-continuous (lsc) at $\x \in \X$ if for every open set $U \subseteq \Y$ such that $F(\x) \cap U \neq \emptyset$ there exists a neighborhood $V$ of $\x$ such that $F(\x') \cap U \neq \emptyset$ for all $\x' \in V$. $F$ is lsc on $\X$ if it is lsc at every point in $\X$.
\label{def:set-valued map lsc}
\end{definition}

\begin{definition}[set-valued map continuity \cite{freeman2008robust}]
    A set-valued map $F: \X \rightsquigarrow \Y$ is continuous on $\X$ if it is lsc and usc on $\X$.
    \label{def:set-valued map continuity}
\end{definition}

\begin{assumption}
    \label{assmp:R set Hausdorff Cont}
    The reachability correspondence $\RSet: \X \rightsquigarrow \X$ is continuous.
\end{assumption}

We consider Markov policies for both agents. 
To initialize the game, at time $t=0$, the Opponent first selects a node $i_0 \in \nodeset$ according to its policy $\sigma_0(\Q, \mathcal{G})$.
Here, we make the policy dependence on $\Q$ and the graph $\mathcal{G}$ explicit.
After observing the Opponent's selection~$i_0$, the Player selects a point $\x_0 \in \nsp{0}{i_0}$ according to the policy $\pi_0(i_0)$.
The Player's (deterministic) policies at time ${t \geq 1}$ is given by $\pi_t(\x_{t-1}, i_{t}) \in \RSet(\x_{t-1}) \cap \nsp{t}{i_{t}}$, which explicitly considers the Opponent-selected convex body $\nsp{t}{i_t}$at time $t$ and the Player's reachability constraint.
The Opponent's policy at time $t \geq 1$ is given by $\sigma_t(\x_{t-1}, i_{t-1}) \in \neighbor{i_{t-1}}$, which reflects the graph constraint on the Opponent's state. 
We collect the sequences of the policies used by the Player and the Opponent to the strategies $\pi = \{\pi_t\}_{t=0}^T$ and $\sigma=\{\sigma_t\}_{t=0}^T$, respectively. 

A strategy pair $(\pi, \sigma)$ provides a trajectory for the Player and, consequently, induces a total movement cost  $C(\pi,\sigma)$ characterized as $\sum_{t=0}^{T-1}{c~(\x_t, \x_{t+1})}$. 
%
%
We are interested in subgame perfect equilibria, namely, 
equilibria where, at each stage of the game, the Player always minimizes its future cumulative cost-to-go while the Opponent maximizes it. 
We denote the total cost under a subgame perfect equilibrium as the optimal total cost $C^*$.

\begin{problem} \label{problem1}
   Given a graph $\mathcal{G}$, a collection of solid and compact convex bodies $\Q=\big\{\nsp{t}{i}\big\}_{t=0, i=1}^{T, |\nodeset|}$, a cost function $c$, and a reachablility correspondence $\RSet$, find  the optimal total cost $C^*$ of the aCBC game along with the corresponding optimal strategies for the Player and the Opponent under the information structure in Assumption~\ref{assmp:info}.
\end{problem}

To solve  Problem~\ref{problem1}, we follow a value-based approach, where at each decision point, the agents compute a policy that optimizes its ``cost-to-go.'' 
The rest of the paper will address the technical details regarding the solution to this
optimization problem.


\subsection{Connection to the Dynamic Defender-Attacker Blotto Game}  
At first glance, 
one may find Assumption~\ref{assmp:feasibility} restrictive.
However, for safety-critical problems, it is common to first construct a set of policies that are safe/feasible and then consider optimality in the safe domain.
One potential application of the aCBC game is the dynamic defender-attacker Blotto game (dDAB)~\cite{shishika2021dynamic}, which is a dynamic and adversarial resource allocation problem in a graph environment. 
In the dDAB a team of defender robots is tasked to ensure a numerical advantage over a team of attackers at every node. 
The two teams reallocate their robots in sequence and each robot (resource) can move at most one hop at each time step. 
The dDAB game is formulated as a game of kind, and the game terminates with the attacker’s victory if any node has more attacker robots than defender robots.
In \cite{shishika2021dynamic} is is shown that the defender's feedback strategy is specified by the safe sets given as a function of the attacker's allocation.
In effect, the attacker (the Opponent) is selecting a sequence of safe sets, which the defender's allocation (the Player's state) must stay in, for the sake of successful defense.
The collection of safe-sets in dDAB satisfies Assumption~\ref{assmp:feasibility} and is one of the major motivations of this work.
Consequently, the aCBC framework can naturally extend the dDAB game to a game-of-degree formulation by introducing costs to the defender movements.

\section{Optimal Value Functions}   \label{sec:value-function}

To reflect the different information available to the two agents at their decision points, we introduce two value functions: $\playerV_t~(\x_{t-1}, i_t)$ for the optimal value function of the Player and $\opponentV_{t}~(\x_{t-1}, i_{t-1})$ for the Opponent at time $t$. 
These two optimal value functions will be computed through a backward induction scheme, similar to other finite horizon decision-making problems~\cite{bertsekas2012dynamic, littman1996algorithms}.

\subsection{Backward Induction}

At the terminal time step $T$, the Player has knowledge on its previous state $\x_{T-1}$ and the Opponent state $i_T$, and the Player is about to make  its final move.
Since there are no moves after time $T$, the Player only needs to consider the optimality with respect to the immediate cost $c~(\x_{T-1}, \x_T)$.
Consequently, the optimal terminal value for the Player can be formulated as 
\begin{equation}       
    \label{eqn:V_T_def}
    \playerV_{T}~(\x_{T-1}, i_{T}) \triangleq \inf_{\x_{T} ~ \in ~ \ic{T}}{ c~(\x_{T-1}, \x_{T}) }.
\end{equation}
In other words, the above value depicts the best feasible outcome for the Player, given its previous state $\x_{T-1}$ and the Opponent state $i_T$ at the terminal time step $T$.

For time steps $t\in \{1, \ldots, T-1\}$, the Player needs to optimize its selection of a new state $\x_t$ in order to minimize both the immediate cost and the worst-case future cost. 
Specifically, the Player has to also consider the fact that the Opponent will observe the  new Player state $\x_t$ and then best-respond with $i_{t+1}$ to maximize the future cumulative costs. 
Consequently, the optimal value for the Player is formulated as 
\begin{equation}
    \label{eqn:V_t_def}
    \playerV_{t}~(\x_{t-1},i_{t}) \triangleq \inf_{\x_{t} ~\in~ \ic{t}}{\left\{c~(\x_{t-1}, \x_{t}) + \max_{i_{t+1}~\in~\neighbor{i_{t}}}\playerV_{t+1}~(\x_{t},i_{t+1})\right\}}.
\end{equation}

Finally, for the initial Player state selection at $t=0$ there is no reachability constraint or immediate cost. 
As a result, the optimal value function only depends on the initial Opponent state $i_0$, while the optimization only covers the worst-case future cost similar to the value function in~\eqref{eqn:V_t_def}, and we have
\begin{equation}
    \label{eqn:V_0_def}
    \playerV_{0}~(i_{0}) \triangleq \inf_{\x_{0} ~\in~ \Q_{0}^{(i_{0})}} \; \max_{i_{1}~\in~\neighbor{i_{0}}}\playerV_{1}~(\x_{0},i_{1}).
\end{equation}

The Opponent's optimal value function is constructed similarly to that of the Player.
The only difference comes from the information structure. 
Namely, at time step $t$, the Opponent makes a decision based on the previous Player state $\x_{t-1}$ and its own previous state $i_{t-1}$.
Formally, the Opponent's optimal values are formulated as
\begin{align}
    \opponentV_{T}~(\x_{T-1}, i_{T-1}) &\triangleq \max_{i_{T}~ \in~ \neighbor{i_{T-1}}}{\left\{\inf_{\x_{T}~\in~\ic{T}}{c~(\x_{T-1}, \x_{T})}\right\}},
    \label{eqn:U_T_def}
    \\
    \opponentV_{t}~(\x_{t-1}, i_{t-1})&\triangleq \max_{i_{t} ~\in~ \neighbor{i_{t-1}}}{\left\{\inf_{\x_{t} ~\in~ \ic{t}}{\{c~(\x_{t-1}, \x_{t})+\opponentV_{t+1}~(\x_{t}, i_t)\}}\right\}},
    \label{eqn:opponent-u-t-sup}
    \\
    \opponentV_{0}~(\mathcal{G}, \Q) &\triangleq \max_{i_{0}~\in~\nodeset} {~\inf_{\x_{0}~\in~\nsp{0}{i_{0}}}} {\opponentV_{1}~(\x_{0}, i_0)}.
    \label{eqn:U_0_def}
\end{align}

\begin{remark}
    The Opponent's value $U_0$ is equivalent to the optimal total cost $C^*$ of the aCBC game.
\end{remark}

\begin{remark}
    From time step $1$ to $T$, we implicitly assume $\x_{t-1} \in \nsp{t-1}{i_{t-1}}$ and $i_{t-1} \in \nodeset$ for all value functions $\opponentV_t~(\x_{t-1}, i_{t-1})$.
    Likewise, for the value functions $\playerV_t~(\x_{t-1}, i_t)$ we assume that $\x_{t-1} \in \nsp{t-1}{i_{t-1}}$ and $i_t \in \neighbor{i_{t-1}}$.

    
    
\end{remark}

\subsection{Continuity of the Optimal Value Functions}

The first question one may ask regarding the value functions is whether the infimum in \eqref{eqn:V_T_def}-\eqref{eqn:U_0_def} can be attained. 
To answer this question, we first show that the intersection correspondence 
$\Theta_t^{(i)}$ is continuous, then we prove the continuity of the value functions with respect to the $\x$-arguments.
%
Assumption \ref{assmp:R set Hausdorff Cont} only regards the continuity of the reachability correspondence $\RSet$ rather than of the intersection correspondence $\Theta_t^i$.
To bridge this gap, we present the following lemma to guarantee the continuity of $\Theta_t^i$.


\begin{restatable}{lemma}{intersectC}
    Let $\Gamma: \X \rightsquigarrow \Y$ be continuous, and let $\Gamma(\x)$ be compact and convex for all $\x \in \X$. Consider a closed convex set $F \subseteq \Y$ such that $\mathrm{int} (\Gamma(\x) \cap F) \neq \emptyset$ for all $\x \in C$, where $C \subseteq \X$ is closed. Then, the correspondence $\Xi: C \rightsquigarrow \Y$ defined by $\Xi(\x) = \Gamma(\x) \cap F$ is continuous on $C$. 
    \label{lmm:intersection-continuity}
\end{restatable}

\begin{proof}
    Please see Appendix~\ref{appdx:intersect_correspondence_cont} for details.
\end{proof}



The following lemma provides insight into the continuity of marginal functions of the form

\begin{equation}
    \label{eqn:marginal-function}
    \phi(\x) = \inf_{\y \in \Gamma(\x)} f(\x, \y)
\end{equation}

\begin{lemma}[Proposition 2.9 in~\cite{freeman2008robust}]
    \label{lmm:marginal-fx-cont}
    Consider a continuous function $f:\X \times \Y \rightarrow \R$ and a continuous correspondence $\Gamma: \X \rightsquigarrow \Y$. If $\Gamma$ has compact values, then the marginal function $\phi: \X \rightarrow \R$ in \eqref{eqn:marginal-function} is continuous. 
\end{lemma}

Note that the value functions in \eqref{eqn:V_t_def} and \eqref{eqn:opponent-u-t-sup} take the form of a marginal function. 
Consequently, we can utilize Lemma~\ref{lmm:marginal-fx-cont} to prove the continuity of the value functions.
This result is stated in the following theorem.

\begin{restatable}{theorem}{vc} \label{thm:value_fx_cont}
    For all $t \in \{1, \ldots, T\}$ and $i \in \nodeset$, $j \in \neighbor{i}$, the optimal value functions $\playerV_t~(\cdot, j): \nsp{t-1}{i} \rightarrow \R$ and $\opponentV_t~(\cdot, i): \nsp{t-1}{i} \rightarrow \R$ are both 
    continuous.
\end{restatable}
\begin{proof}
    See Appendix~\ref{appdx:lemma_value_uniform_cont} for details.
\end{proof}
\begin{remark}
    For all $t=1,\ldots,T$, and for all $i_{t-1}$, $i_t \in \nodeset$ such that $i_t \in \neighbor{i_{t-1}}$, the infima  in the expressions of $\playerV_t$ in \eqref{eqn:V_T_def}-\eqref{eqn:V_0_def} are attainable and finite.
    \label{rmk:UV extrema attainable and finite}
    
    
    
\end{remark}



Owing to Remark~\ref{rmk:UV extrema attainable and finite}, we can replace the infimum in the definitions of the value functions with the  minimum. 
The resulting optimal value functions of the Player can therefore be re-written as
\begin{subequations}
    \label{eqn:player_V}
    \begin{align}
        \playerV_T~(\x_{T-1}, i_T) &= \min_{\x_T~\in~\ic{T}} c~(\x_{T-1}, \x_T),
        \label{eqn:playerV_T}
        \\
         \playerV_t~(\x_{t-1}, i_t) &= \min_{\x_t ~\in ~\ic{t}} \left\{c~(\x_{t-1}, \x_t) + \max_{i_{t+1} ~\in~ \neighbor{i_t}} \playerV_{t+1}~(\x_t, i_{t+1})\right\},
         \quad \forall~t \in \{1,\ldots,T-1\},~
        \label{eqn:playerV_t}
        \\
        \playerV_0~(i_0) &= \min_{\x_0 ~\in ~\nsp{0}{i_0}} ~ \max_{i_1 ~\in ~\neighbor{i_0}} \playerV_1~(\x_0, i_1),
        \label{eqn:playerV_0}
    \end{align}
\end{subequations}

Similarly, the optimal value functions of the Opponent are given as follows.
\begin{subequations}
    \label{eqn:opponent_V}
    \begin{align}
        \opponentV_T~(\x_{T-1}, i_{T-1}) &= \max_{i_T ~ \in ~ \neighbor{i_{T-1}}} \left\{\min_{\x_T ~ \in ~ \ic{T}} c~(\x_{T-1}, \x_T)\right\},
        \label{eqn:opponentV_T}\\
         ~\opponentV_t~(\x_{t-1}, i_{t-1}) &= \max_{i_t ~\in ~\neighbor{i_{t-1}}} \left\{\min_{\x_t ~\in ~ \ic{t}} \{c~(\x_{t-1}, \x_t) + \opponentV_{t+1}~(\x_t, i_t)\}\right\},
         ~~ \forall~t \in \{1,\ldots, T-1\},
        \label{eqn:opponentV_t} \\
        \opponentV_0~(\mathcal{G}, \Q) &= \max_{i_0 ~\in~ \nodeset} ~\min_{\x_0 ~\in~ \nsp{0}{i_0}} \opponentV_1~(\x_0, i_0).
        \label{eqn:opponentV_0}
    \end{align}
\end{subequations}

\subsection{Relationship between the two Optimal Value Functions}

It should be clear from \eqref{eqn:player_V} and \eqref{eqn:opponent_V} that the two value functions $\opponentV_t$ and $\playerV_t$ are related. 
Their relation is formalized in the following lemma.
\begin{restatable}{lemma}{UV}
    \label{lmm:U=max_V}
    For all $t \in \{1, \ldots, T\}$, the Opponent value $\opponentV_t~(\x_{t-1}, i_{t-1})$ is related to the Player value $\playerV_t~(\x_{t-1}, i_t)$ via 
    \begin{equation}
        \opponentV_t~(\x_{t-1},i_{t-1}) = \max_{i_t ~\in~ \neighbor{i_{t-1}}} \playerV_t~(\x_{t-1}, i_t).
        \label{eqn:Ut=max Vt}
    \end{equation} 
    Similarly, for all $t \in \{1, \ldots, T-1\}$, the Player value $\playerV_t~(\x_{t-1}, i_t)$ is related to the Opponent value $\opponentV_{t+1}~(\x_t, i_t)$ via
    \begin{equation}
        \playerV_t~(\x_{t-1}, i_t) = \min_{\x_t ~\in~ \ic{t}} \{c~(\x_{t-1}, \x_t) + \opponentV_{t+1}~(\x_t, i_t)\}.
        \label{eqn:Vt = min (c + Ut+1)}
    \end{equation}
    Furthermore, for $t = 0$,
    \begin{equation}
        \label{eqn:U0 in terms of V0}
        \opponentV_0~(\mathcal{G}, \Q) = \max_{i_0~\in~\nodeset} \playerV_0~(i_0),
    \end{equation}
    \begin{equation}
    \label{eqn:V0 in terms of U1}
        \playerV_0~(i_0) = \min_{\x_0 ~\in \nsp{0}{i_0}} \opponentV_1~(\x_0, i_0).
    \end{equation}
\end{restatable}

\begin{proof}
    See Appendix~\ref{appdx:UV-relation}.
\end{proof}


\subsection{Optimal Policies} \label{subsec:optimal-policy}

With the optimal value functions computed, Remark~\ref{rmk:UV extrema attainable and finite} can be used to obtain the optimal policies of the Player and the Opponent.
Specifically, 
for the Player the optimal policy can be obtained as follows
\begin{subequations}
\label{eqn:player-optimal-policy}
    \begin{align}
        \pi_T^*~(\x_{T-1}, i_T) &\in \argmin_{\x_T \; \in \; \ic{T}} c~(\x_{T-1}, \x_{T}), \label{eqn:player-optimal-policy-T}\\
        \pi_t^*~(\x_{t-1}, i_t) &\in \argmin_{\x_t \;\in\; \ic{t}} \{c~(\x_{t-1}, \x_t) + \opponentV_{t+1}~(\x_t, i_t)\}, \quad \forall \; t=1,\ldots,T-1, \label{eqn:player-optimal-policy-t}\\
        \pi_0^*~(i_0) &\in \argmin_{\x_0 \; \in \; \nsp{0}{i_0}} \opponentV_1~(\x_0, i_0). \label{eqn:player-optimal-policy-0}
    \end{align}
\end{subequations}

Similarly, for the Opponent the optimal policy can be obtained as follows
\begin{subequations}
\label{eqn:opponent-optimal-policy}
    \begin{align}
    \sigma_t^*~(\x_{t-1},i_{t-1}) &\in \argmax_{i_t ~\in~ \neighbor{i_{t-1}}} \playerV_t~(\x_{t-1}, i_t), \quad \forall \; t=1,\ldots,T, \label{eqn:opponent-optimal-policy-t}\\
    \sigma_0^*~(\mathcal{G}, \Q) &\in \argmax_{i_0~\in~\nodeset} \playerV_0~(i_0). \label{eqn:opponent-optimal-policy-0}
\end{align}

Consequently, given the Player state $\x$ and Opponent state $i$, the optimal policies can be obtained using backward propagation.
However, since the $\x$-argument of the value function is taken in an uncountable set, storing and optimizing the value functions is challenging.
One natural approach is to properly discretize the domain of the $\x$-argument and approximate $\playerV_t$ and $\opponentV_t$ with their values at the vertices of a mesh in the $\x$-domain.
%
Section~\ref{sec:algorithm} will develop an algorithm that implements this discretization idea.


\end{subequations}


\section{Algorithmic Solution} \label{sec:algorithm}

To numerically compute the optimal values, we propose an algorithm that meshes the domain $\X$ and the approximate $U_t$ and $V_t$ at the vertices of the mesh, similar to the approaches discussed in~\cite{rao2009survey}.
In order to derive approximation error bounds induced by the discretization,
we first strengthen the continuity properties on $U_t$ and $V_t$. 

\subsection{Lipschitz Continuity of the Value Functions}
\label{subsec:value function Lipschitz Continuity}

We impose the following two assumptions to ensure the Lipschitz continuity of $\playerV_t$ and $\opponentV_t$.


\begin{restatable}{assumption}{CLC}   \label{assmp:cost-L-cont}
The cost function $c: \X \times \X \to \mathbb{R}$ is Lipschitz continuous under the Manhattan distance.
That is, for all $(\x, \y), (\x', \y') \in \X \times \X$,
    \begin{equation*}
        \abs{c~(\x, \y) - c~(\x', \y')} \leq L_c \, (\norm{\x-\x'} + \norm{\y-\y'}),
    \end{equation*}
    where $L_c$ denotes the Lipschitz constant.


\end{restatable}




Before introducing the next assumption, we first need to define the \textit{Hausdorff distance} between two sets.

\begin{definition}[Hausdorff distance~\cite{freeman2008robust}]
    \label{def:Hausdorff distance def}
    Given two subsets $E$ and $F$ of the normed space $\X$, the Hausdorff distance $\distH{E,F}$ between $E$ and $F$ is defined as
    \begin{equation*}
        \displaystyle\distH{E, F} = 
       \max \big\{\hspace{-2pt}\sup_{\x \in E} \inf_{\y \in F} \norm{\x - \y}\hspace{-1pt},  \sup_{\y \in F} \inf_{\x \in E} \norm{\x - \y}\hspace{-2pt}\big\}.
    \end{equation*}
\end{definition}
In case the sets $E$ and $F$ are compact, the sup and inf can be replaced by max and min, respectively, 
\begin{restatable}{assumption}{RLC}   \label{assmp:R-set-L-cont}
    The correspondence $\Theta_t^{(i)}$ is uniformly $L_\Theta$-Lipschitz continuous with respect to $i \in \nodeset$ and $t \in \{1, \ldots, T\}$ under the Hausdorff distance.
    %
    That is, there exists a constant $L_\Theta$  such that, for all $i \in \nodeset$ and $t \in \{1, \ldots, T\}$, the following holds:
    \begin{equation*}
        \distH{\Theta_t^{(i)}(\x), \Theta_t^{(i)}(\x')} \leq L_\Theta \, \norm{\x - \x'}, 
        \quad \forall \; \x, \x' \in \nsp{t-1}{j}, \text{ where } i \in \neighbor{j}.
    \end{equation*} 
\end{restatable}


\begin{restatable}{theorem}{vlc}
    \label{thm:value_fx_L_cont}
    Under Assumptions~\ref{assmp:cost-L-cont} and~\ref{assmp:R-set-L-cont}, for all $t \in \{1, \ldots, T\}$, $i_{t-1} \in \nodeset$ and $i_t \in \neighbor{i_{t-1}}$, the optimal value functions $\playerV_t~(\cdot, i_t)$ and $\opponentV_t~(\cdot, i_{t-1})$ are both $L_{v,t}$-Lipschitz continuous on $\nsp{t-1}{i_{t-1}}$ with 
    Lipschitz constant is given by 
    \begin{align}
    \label{eqn:L_v_t expr}
        L_{v,t} = L_c \, \sum_{k=1}^{T-t+1} (1+L_\Theta)^k.
    \end{align}
    That is, for all $\x_{t-1}$, $\x'_{t-1} \in \nsp{t-1}{i_{t-1}}$,
    \begin{equation*}
        \abs{\playerV_t~(\x_{t-1}, i_t) - \playerV_t~(\x'_{t-1}, i_t)} \leq L_{v,t} \norm{\x_{t-1} - \x'_{t-1}},
    \end{equation*}
    and 
    \begin{equation*}
        \abs{\opponentV_t~(\x_{t-1}, i_{t-1}) - \opponentV_t~(\x'_{t-1}, i_{t-1})} \leq L_{v,t} \norm{\x_{t-1} - \x'_{t-1}}.
    \end{equation*}
    \end{restatable}
\begin{proof}
See Appendix~\ref{appdx:lemma_value_L_cont} for details.
\end{proof}

\subsection{Discretization Scheme and Discretized Value Functions}
\label{subsec:discretization-scheme-and-discretized-value-functions}

Theorem~\ref{thm:value_fx_L_cont} implies that the Lipschitz constant $L_{v,t}$ decreases monotonically as the time step $t$ approaches the horizon~$T$.
Naturally, finer discretization resolution is preferred at the beginning of the game to ensure low approximation error. 
Consequently, we allow different resolutions at different time steps. 
We use $\delta_{\X,t}$ to denote the discretization size of the state space $\X$ at time step $t$, and we denote the set of vertices on the mesh at time $t$ by $\widehat{\X}_t = \{\widehat{\x}_t^{k}\}_{k=1}^{M_t}$.
To ensure that the $\x$-optimization domain $\ic{t}$ in~\eqref{eqn:player_V} is properly discretized with a discretization size $\delta_{\X,t}$, we require 

\begin{subequations}
    \label{eqn:discretization-size}
    \begin{align}
        \min_{\widehat{\x}_t \, \in \, \widehat{\X}_t} \norm{\widehat{\x}_t - \x_t} \leq \delta_{\X,t}, \quad &\forall \; \x_t \in \X,~~  t\in\{,\ldots,T\}, \label{eqn:discretization-X}\\
      \min_{\widehat{\x}_0 \, \in \, \widehat{\X}_0 \, \cap \, \nsp{0}{i_0}} \norm{\widehat{\x}_0 - \x_0} \leq \delta_{\X, 0}, \quad & \forall \; \x_0 \in \nsp{0}{i_0},~~ i_0 \in \nodeset, \label{eqn:discretization-intersection-0}\\
        \quad \min_{\widehat{\x}_t \, \in \, \widehat{\X}_t \, \cap \, \widehatic{t}} \norm{\widehatx_t - \x_t} \leq \delta_{\X,t}, \quad &\forall \; \x_t \in \widehatic{t}, ~~  t \in \{1, \ldots, T\},\label{eqn:discretization-intersection}\\
        & ~~ \;\widehatx_{t-1} \in \widehatX_{t-1} \cap \nsp{t-1}{j} \text{ where } i_t \in \neighbor{j}. \nonumber 
    \end{align}
\end{subequations}

In the above criteria, we first require that the mesh has the required resolution over the whole domain $\X$ at all time steps as in~\eqref{eqn:discretization-X}.

\begin{figure}[!htb]
    \centering
    \begin{minipage}{0.48\textwidth}
        \centering
        \includegraphics[width=0.9\linewidth]{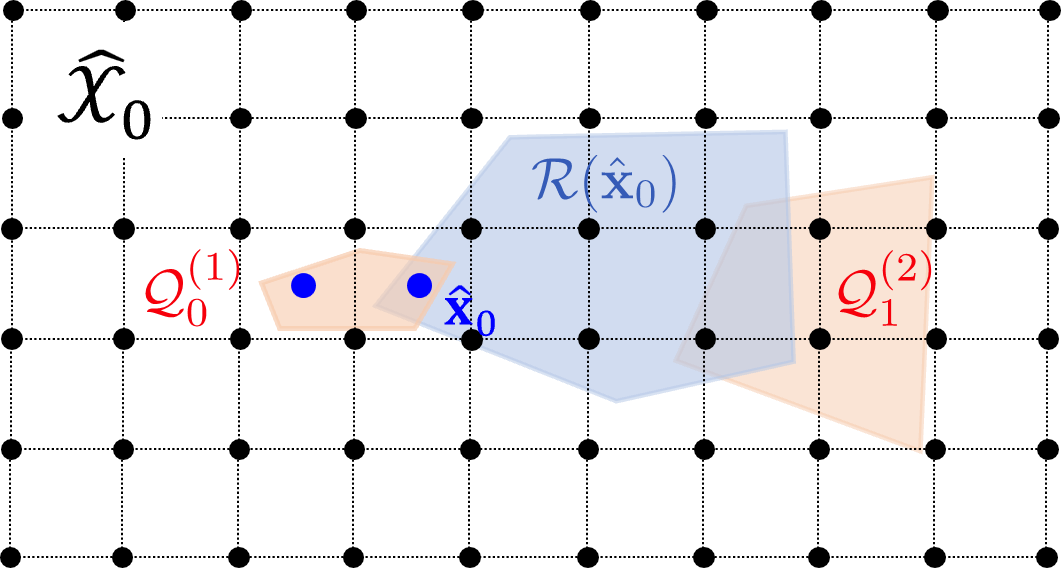}
    \end{minipage}
    \begin{minipage}{0.48\textwidth}
        \centering\includegraphics[width=0.9\linewidth]{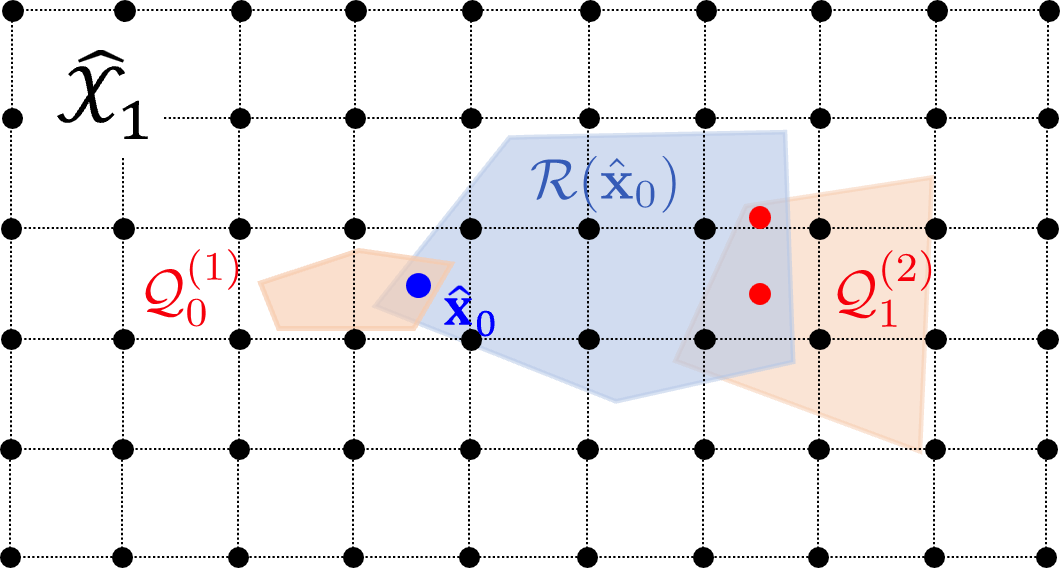}
    \end{minipage}
    \caption{A demonstration of mesh construction at time steps $t=0$ and $t=1$.}
    \label{fig:discretization-demo}
\end{figure}

An example of the discretization scheme is presented in \figref{fig:discretization-demo}.
The black vertices are constructed to discretize the domain $\X$ based on~\eqref{eqn:discretization-X}.
Note that there is no black vertex in $\nsp{0}{1}$. 
Consequently, two extra blue vertices are added to $\widehatX_{0}$ to satisfy~\eqref{eqn:discretization-intersection-0}.
Furthermore, although the intersection $\RSet(\widehat{\x}_0) \cap \nsp{1}{2}$ contains a black vertex from $\widehatX_1$, the mesh is not fine enough within the intersection to satisfy condition \eqref{eqn:discretization-intersection}. 
As a result, two red vertices are added to $\widehatX_1$.

Since the discretized optimization domain at time $t$ is $\widehat{\X}_t \cap \icjx{t}{i_t}{\widehat{\x}_{t-1}}$, one also needs to ensure that the mesh is fine enough in $\icjx{t}{i_t}{\widehat{\x}_{t-1}}$, and hence the additional requirements in~\eqref{eqn:discretization-intersection-0} and~\eqref{eqn:discretization-intersection} are added.

With the discretization of $\X$ at every time step, the Player restricts its action selection $\x_t$ at time step $t$ to the vertices in $\widehat{\X}_t$, which leads to a decrease in  performance.
We consider the worst-case scenario, where the Opponent knows the mesh used by the Player.
The optimization domains of the resulting discretized Player value functions are then represented using the vertices of the mesh $\widehatX_t$.
We denote the discretized value functions as $\widehat{\playerV}$ and $\widehat{\opponentV}$ for the Player and the Opponent, respectively.
For example, the Player's value function at time step $t$ between $1$ and $T-1$ is given by
\begin{equation}
    \label{eqn:apprx-value-example}
    \widehat{\playerV}_t~(\widehat{\x}_{t-1}, i_t) = \min_{\widehat{\x}_t \, \in \,\widehat{\X}_t \,\cap\, \widehatic{t}} \left\{c~(\widehat{\x}_{t-1}, \widehat{\x}_t) + \max_{i_{t+1} ~\in~ \neighbor{i_t}} \widehat{\playerV}_{t+1}~(\widehat{\x}_t, i_{t+1})\right\}.
\end{equation}
For the detailed definition of all discretized value functions, see Appendix~\ref{appdx:discretized-value}.


We denote the optimal strategies induced from the discretized value functions as 
$\widehat{\pi}^*$ and $\widehat{\sigma}^*$ for the Player and the Opponent, respectively. 
%
Specifically, the Player's discretized optimal policy at time step $t$ between $1$ and $T-1$ is given by
\begin{equation}
    \label{eqn:apprx-opt-policy-example}
    \widehat{\pi}_t^*~(\widehatx_{t-1}, i_t) \in \argmin_{\widehatx_t \, \in \, \widehatX_t \, \cap \, \widehatic{t}} \{c\halfspace(\widehatx_{t-1}, \widehatx_t) + \widehat{\opponentV}_{t+1}\halfspace(\widehatx_t, i_t)\}.
\end{equation}

For the detailed definition of discretized optimal policies, see Appendix~\ref{appdx:discretized-policy}.

\begin{remark}
    At every time step $t$, the discretized value $\widehat{\playerV}_t$ corresponds to the optimal worst-case performance of the Player using the discretization scheme $\widehatX$.
    This implies that the Opponent has perfect knowledge of the discretization scheme used by the Player. 
    Moreover, the Opponent exploits this knowledge when maximizing the Player cost-to-go at every time step $t$.
\end{remark}

If both the Player and Opponent apply strategies $\widehat{\pi}^*$ and $\widehat{\sigma}^*$, then the game value $\widehat{\opponentV}_0~(\mathcal{G},\Q)$ in \eqref{eqn:apprx-opponentV_0} is realized.
In this case,  $\widehat{\opponentV}_0~(\mathcal{G},\Q)$ denotes the game total cost approximated by discretization when both agents apply their optimal strategies.
On the other hand, if the Opponent unilaterally deviates and applies a strategy different from $\widehat{\sigma}^*$, then the game will terminate with a total cost less than or equal to $\widehat{\opponentV}_0~(\mathcal{G}, \Q)$
, which is favorable to the Player. 
Similarly, the total cost will be greater than or equal to $\widehat{\opponentV}_0~(\mathcal{G}, \Q)$ if the Player unilaterally deviates from $\widehat{\pi}^*$, putting the Player at  a disadvantage.
Whenever one agent deviates from its optimal strategy, the optimal strategy of the undeviated agent also changes accordingly.
Moreover, it is also noteworthy that $\widehat{U}_0(\mathcal{G},\Q) \geq U_0(\mathcal{G},\Q)$ as it is shown in the next section, 
which means that the approximate equilibrium induced by discretizing the {Player's} domain yields, as expected, a lower Player performance than the actual equilibrium of the game.


\subsection{Error Bounds for Discretization} 
In this subsection, we discuss the relation between the discretization size $\delta_{\X, t}$ and the performance level $\varepsilon$ that bounds the discretization error $\vert \widehat{\opponentV}_0 ~(\mathcal{G}, \Q) - \opponentV_0~(\mathcal{G}, \Q) \vert$.
%
In the end, we introduce an algorithm that computes the discretized optimal policies.

Based on the Lipschitz continuity properties of the value functions, one expects to have a better approximation of the optimal value function with a finer mesh.
However, the discretized value function at time step $t$ is computed based on the \textit{discretized} value at $t+1$ as shown in~\eqref{eqn:apprx-value-example}.
Consequently, the approximation error propagates over time, and it is relatively unclear how large a discretization size $\delta_{\X, t}$ needs to be at every time step $t$ 
to provide the desired level of performance guarantee.
We answer the above question in the following theorem. 

\begin{restatable}{theorem}{appxv}
\label{thm:discretization-error-bounds}
    Given a discretization scheme $\{\delta_{\X,t}\}_{t=0}^T$ satisfying \eqref{eqn:discretization-X}-\eqref{eqn:discretization-intersection}, the difference between the discretized value function and the optimal value function is bounded, 
    for all $t \in \{1,\ldots, T\}$ and ${\widehat{\x}_{t-1} \in \widehat{\X}_{t-1} \cap \nsp{t-1}{i_{t-1}}}$,
    as follows:
    \begin{align}
        \opponentV_t~(\widehat{\x}_{t-1}, i_{t-1}) \leq \widehat{\opponentV}_t~(\widehat{\x}_{t-1}, i_{t-1}) &\leq \opponentV_t~(\widehat{\x}_{t-1}, i_{t-1}) + L_c \delta_{\X, T} + \sum_{\tau= t}^{T-1} (L_c+L_{v,\tau+1})  \delta_{\X,\tau}, \label{eqn:Ut and hatUt inequality}\\
        \playerV_t~(\widehat{\x}_{t-1}, i_t) \leq  \widehat{\playerV}_t~(\widehat{\x}_{t-1}, i_t) &\leq \playerV_t~(\widehat{\x}_{t-1}, i_t) + L_c  \delta_{\X, T} + \sum_{\tau= t}^{T-1} (L_c+L_{v,\tau+1}) \delta_{\X,\tau} ,\quad \forall \, i_t \in \neighbor{i_{t-1}}.
        \label{eqn:Vt and hatVt inequality}
    \end{align}
\end{restatable}
\begin{proof}
    See Appendix~\ref{appdx:approximation-theorem}.
\end{proof}

\begin{remark}
    \label{remk:discretized-player-performance-drop}
    Theorem~\ref{thm:discretization-error-bounds} states that discretization introduces a drop in Player's performance, assuming that the Opponent properly counteracts. 
    Nonetheless, with a sufficiently fine mesh $\{\delta_{\X, \tau}\}_{\tau=t}^{T}$, the max-min performance $\widehat{\opponentV}_t$ under discretization does not deviate too much from the optimal max-min performance~$U_t$.
\end{remark}

As a direct consequence of Theorem~\ref{thm:discretization-error-bounds}, the following corollary provides error bounds for the game value after discretization. 
In the following, we denote the discretization scheme $\{\delta_{\X, t}\}_{t=0}^T$ as $\delta_\X$.

\begin{restatable}{corollary}{appxvi}
    \label{cor:U0 bound corollary}
    The optimal game value $\widehat{\opponentV}_0~(\nodeset,\Q)$ due to discretization exceeds the optimal game value $\opponentV_0~(\nodeset,\Q)$ by at most 
    \begin{equation}
    \label{eqn:performance-bound}
        \varepsilon (\delta_{\X}) = \sum_{\tau= 1}^{T-1} (L_c+L_{v,\tau+1})  \delta_{\X,\tau} + L_{v,1} \delta_{\X,0} + L_c \delta_{\X, T}.
    \end{equation}
\end{restatable}

\begin{proof}
    See Appendix~\ref{appdx:approximation-theorem}.
\end{proof}
%
%
%
Corollary~\ref{cor:U0 bound corollary} implies that with a proper discretization scheme $\delta_{\X}$, the Player's performance computed using the discretized value functions decreases by at most $\varepsilon (\delta_{\X})$ compared to the optimal performance $\opponentV_0~(\nodeset, \Q)$.
Furthermore, the performance drop diminishes as the discretization sizes $\delta_{\X,t}$ approaches zero.
%
%
Given a desired performance bound $\varepsilon$, one discretization scheme that achieves the desired performance is given by
\begin{equation}
    \label{eqn:desired-discretization-size}
    \deltaX{0} = \frac{\varepsilon}{(T+1)  L_{v,1}}, \quad
    \deltaX{T} = \frac{\varepsilon}{(T+1)  L_{c}}, \quad
    \deltaX{t} = \frac{\varepsilon}{(T+1)  (L_c + L_{v,t+1})},
    \quad
    ~t \in \{1, \ldots, T-1\}.
\end{equation}


The following algorithm summarizes the procedure for computing the discretized optimal values.
Based on the discretized values, the discretized optimal policies can be easily constructed  via~\eqref{eqn:apprx-player-optimal-policy} and~\eqref{eqn:apprx-opponent-optimal-policy}.

\begin{algorithm}[ht]
\SetAlgoLined
\SetKwInput{KwInputs}{Inputs}
\KwInputs{An aCBC instance $\langle \mathcal{G}, \Q, c, \RSet, T \rangle$, desired suboptimality bound $\varepsilon$;}
Compute the Lipscthiz constants $L_c$ and $L_\Theta$\;
Compute the discretization scheme $\delta_\X$ via~\eqref{eqn:desired-discretization-size}\;
Construct meshes $\widehatX$ according to the computed $\delta_\X$ and \eqref{eqn:discretization-size}\;
Compute the discretized optimal value functions under $\widehatX$ according to~\eqref{eqn:apprx-value-example}\;
 \textbf{Return} Discretized value functions $\widehat{V}$ and $\widehat{U}$
 \caption{Solve Discretized Value Function}
 \label{alg:solver}
\end{algorithm}

\section{Numerical Simulations}
\label{sec:application}

For the sake of simplicity, and for illustrative purposes,
we consider an aCBC game on a two-dimensional space where all 
sets are box sets. 
For some real numbers $a_1 < b_1$ and $a_2 < b_2$, we define the two-dimensional box set as the Cartesian product $[a_1, b_1] \times [a_2, b_2]$.
With $\X$ also being a box in $\R^2$, the reachability correspondence is defined  as:
\begin{equation}      \label{eqn:R(x) in Numerical Simulation}
    \RSet(\x) = \{\y \in \X :  \norm{\x - \y}_\infty \leq \rho \},
\end{equation}
for some $\rho > 0$. 
%
Taking the Euclidean norm $c~(\x,\y) = \|\x - \y\|_2$ as the cost function, one can verify that the cost function~$c$ is 1-Lipschitz under the Manhattan distance. 
Furthermore, for all time steps $t$ and for all $i_t \in \nodeset$, the correspondance
$\Theta_t^{(i_t)}$
is 1-Lipschitz continuous under the Hausdorff distance.
The proofs of these facts are provided in Lemma~\ref{lmm:Euclidean cost Lipschitz} and 
Lemma~\ref{lmm:Theta_t 1-Lip in numerical sim} in Appendix~\ref{appdx:proofs for numerical simulation}.
%
Given these Lipschitz constants, one can use~\eqref{eqn:desired-discretization-size} to derive the desired discretization sizes $\delta_\X$ for a given suboptimality bound~$\varepsilon$. 

\begin{figure}[!htb]
    \vspace{-13pt}
    \centering
    \includegraphics[width=0.6\linewidth]{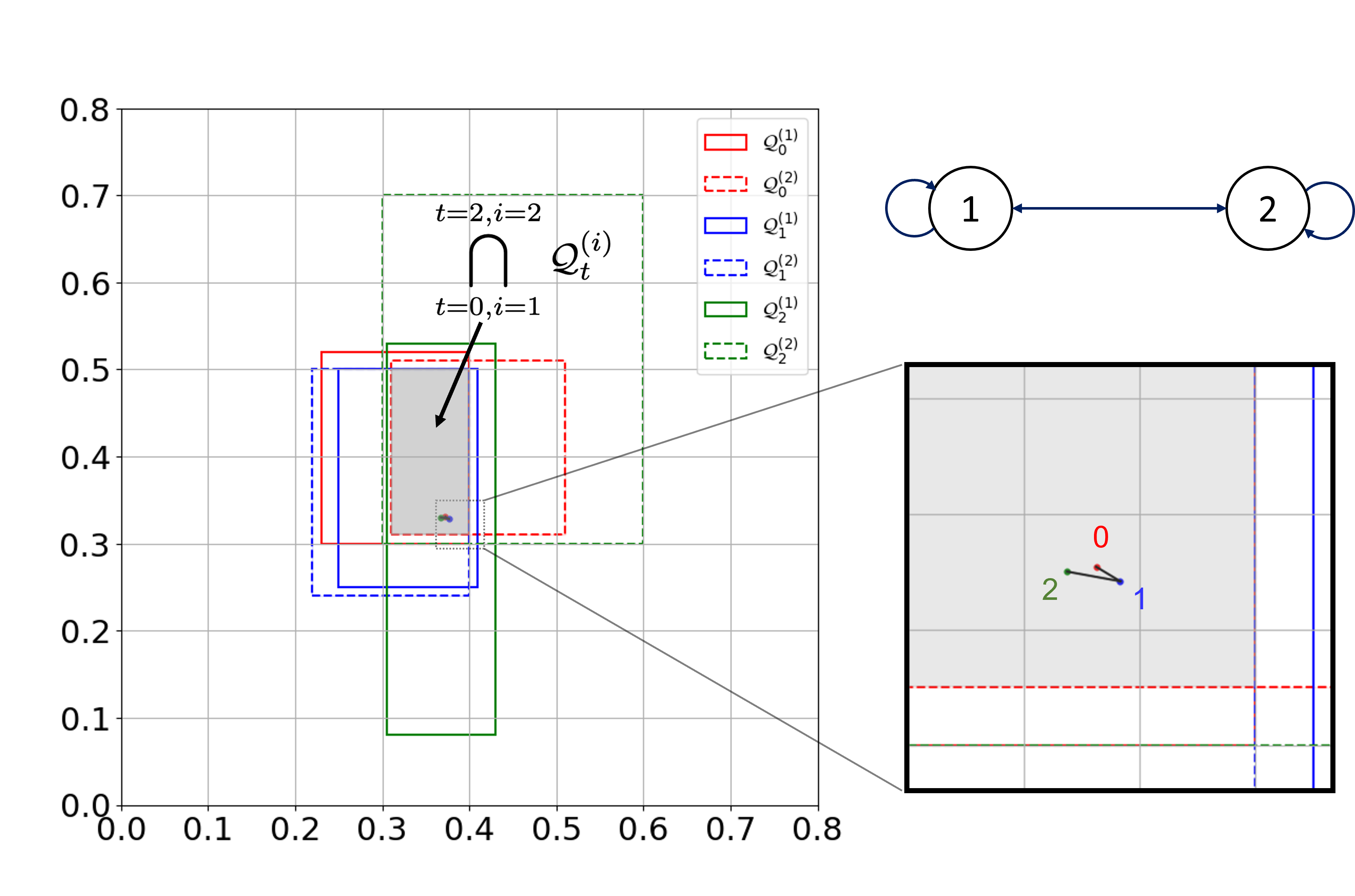}
    \vspace{-10pt}
    \caption{An illustrative aCBC instance with nested convex regions. 
    The left schematic gives the convex sets used for this problem definition. 
    The bottom right schematic shows the optimal policy found using discretization. 
    The top right schematic shows the 2-node graph used in this scenario.}
    \label{fig:nested-acbc}
\end{figure}

We first verify that the discretized algorithm indeed converges to the optimal solution. 
The ``nested'' convex region example in~\figref{fig:nested-acbc} has a simple optimal solution, where the Player starts at any point within the intersection of all convex regions (marked in grey) and does not move. 
Under this optimal strategy, regardless of the actions of the Opponent, the Player can achieve zero cost.

\begin{figure}[h]
    \centering
    \vspace{-5pt}
    \includegraphics[width=0.5\linewidth]{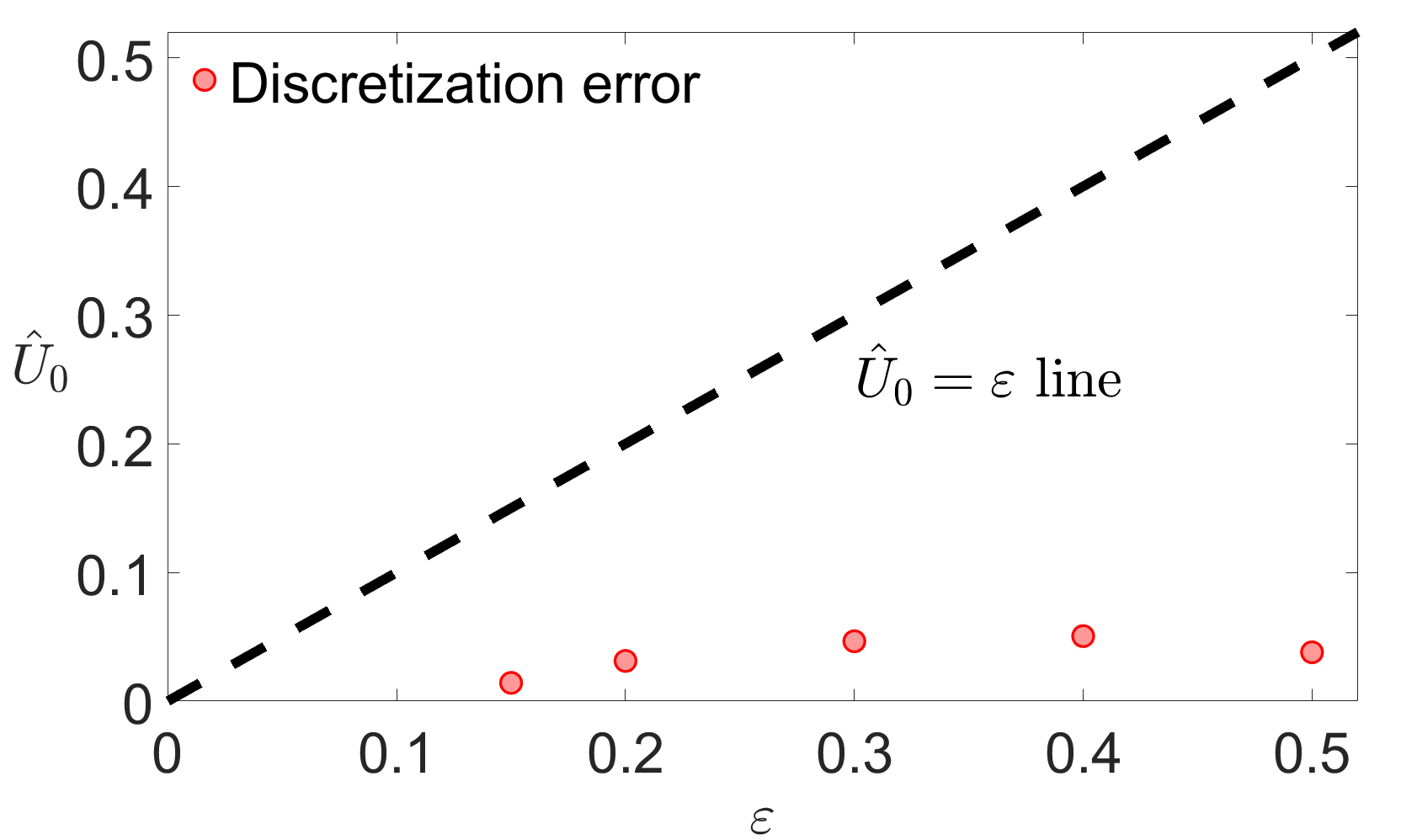}
    \vspace{-5pt}
    \caption{suboptimality bound vs. actual error plot.}
    \vspace{-5pt}
    \label{fig:error-plot}
\end{figure}

However, as the discretization mesh $\widehatX_t$ changes over time, there may not be a point that is a vertex for all intermediate meshes. 
Consequently, the Player may move slightly from time to time under the computed optimal \textit{discretized} policy, incurring
a discretization error (see the zoom-in plot in~\figref{fig:nested-acbc}). 
Since the optimal value $\opponentV_0$ is zero, the discretized value $\widehat{U}_0$ is exactly the discretization error. 
%
%
To verify that the discretization scheme in~\eqref{eqn:desired-discretization-size} indeed achieves the required performance, we run Algorithm~\ref{alg:solver} with different $\varepsilon$ values and plot the corresponding discretization errors in \figref{fig:error-plot}.
One can see that the discretization error diminishes as $\varepsilon$ approaches zero.
Furthermore, the discretization error is bounded by the desired error bound $\varepsilon$ provided in Algorithm~\ref{alg:solver}, which validates the bounds derived in~\eqref{eqn:performance-bound}.


%

\begin{figure}[!h]
    \centering    \includegraphics[width=0.5\linewidth]{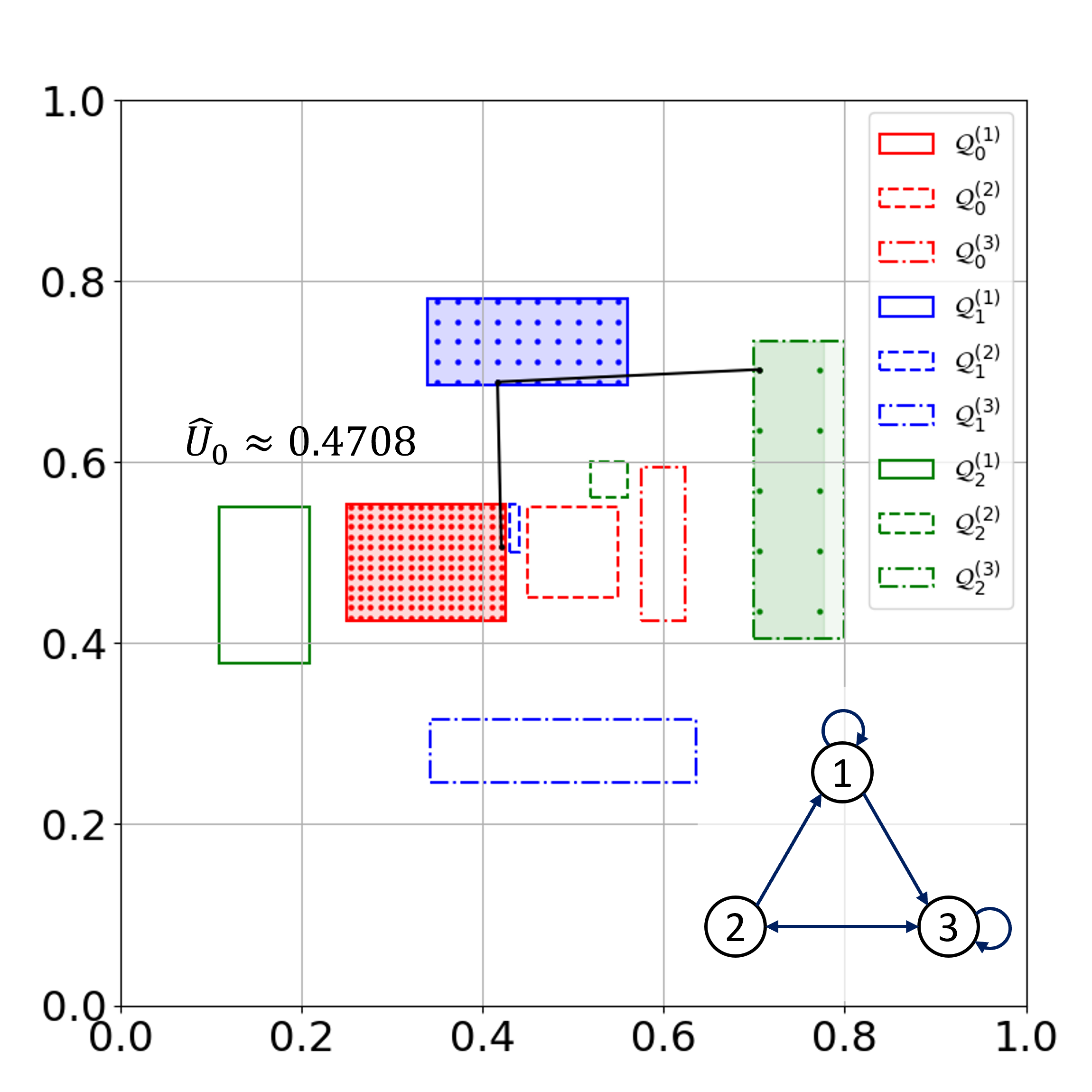}
    \caption{The Max-Min trajectory from the discretized optimal policies. }
    \label{fig:max-min-sol}
\end{figure}

\begin{figure}[!htb]
    \centering
    \begin{minipage}{0.4\textwidth}
    \centering
    \includegraphics[width=0.8\linewidth]{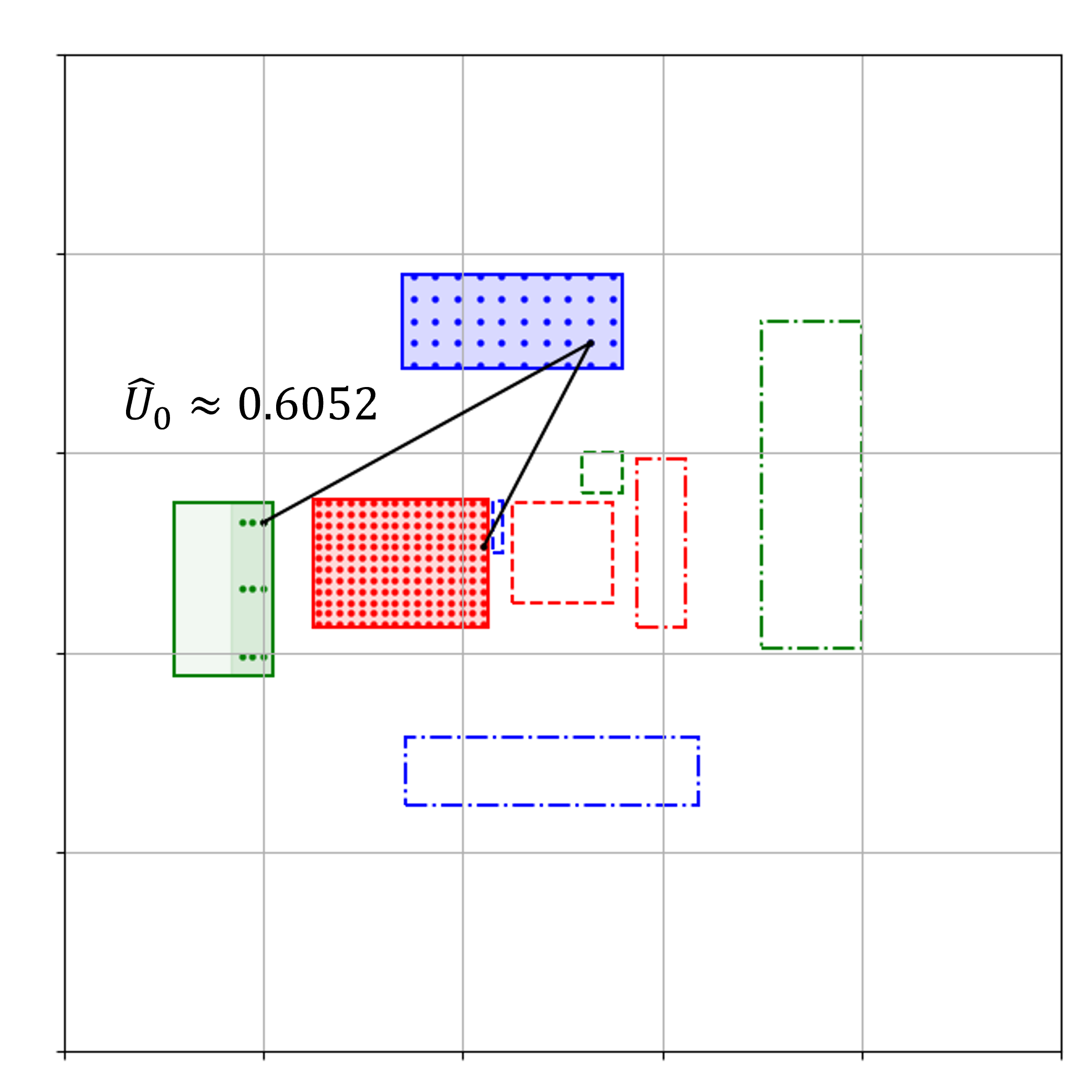}
    \end{minipage}
    \begin{minipage}{0.4\textwidth}
    \centering
    \includegraphics[width=0.8\linewidth]{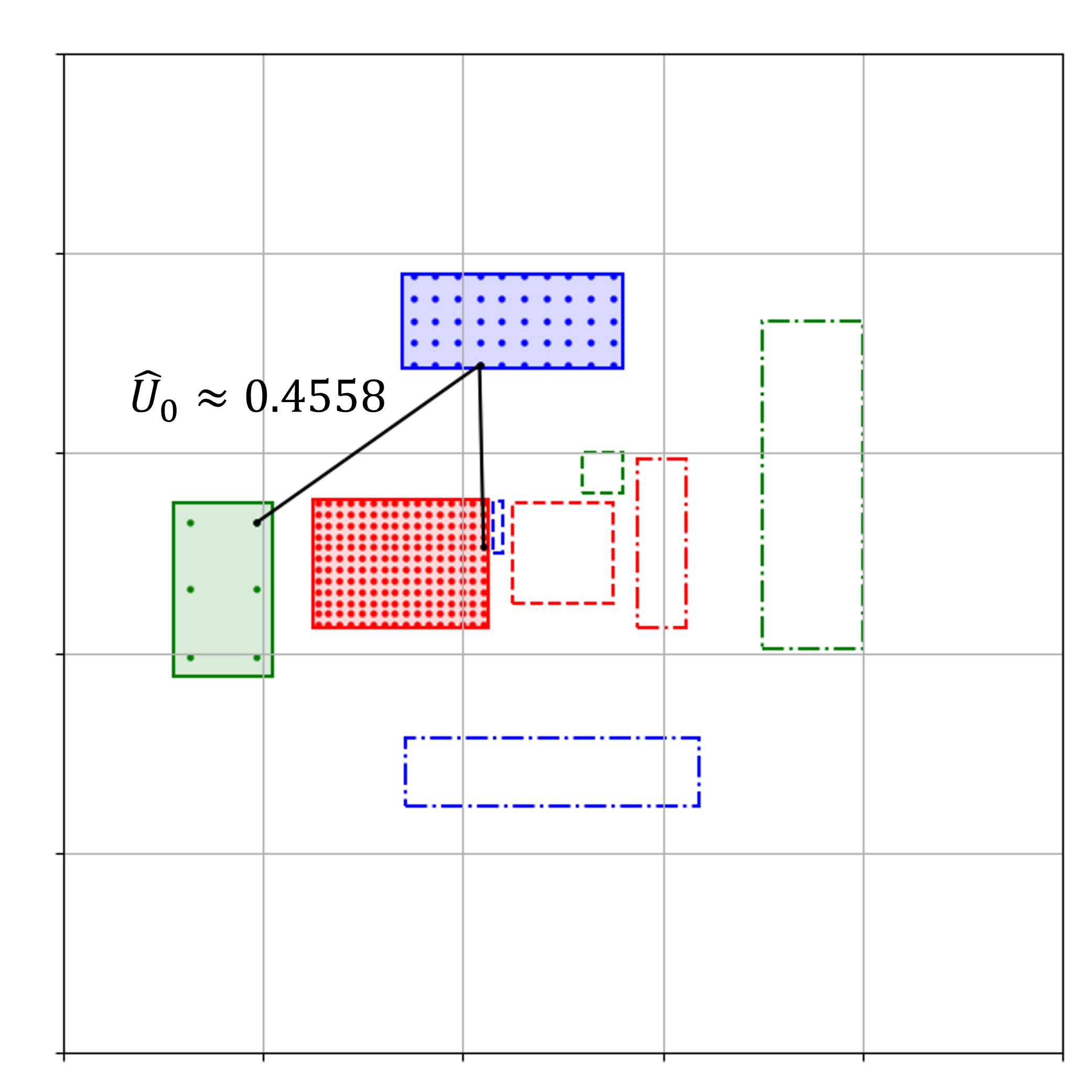}
    \end{minipage}
    \caption{Trajectories with unilateral deviation. The left figures shows the case when the Player deviates and the right figure shows the case when the Opponent deviates.}
    \label{fig:unilateral-deviation}
\end{figure}

Next, we present a more complicated scenario with a graph of three nodes and a two-step horizon, with  reachable sets of size $\rho=0.36$. 
\figref{fig:max-min-sol} shows the convex regions~$\Q$ and the graph $\mathcal{G}$.
If a convex region is selected by the Opponent, it is filled with color. 
The darker color marks the intersection of the reachable sets and the corresponding convex sets.
For example, the darker region in $\nsp{2}{3}$ depicts $\RSet(\x_1) \cap \nsp{2}{3}$, where $\x_1$ is the point selected in $\nsp{1}{1}$.
%
%
The trajectory (in black color) in \figref{fig:max-min-sol} is the max-min trajectory induced by the discretized optimal policies computed under a desired error bound $\varepsilon = 0.2$.
The trajectory of the Opponent is $1 \to 1 \to 3$.
One can see that at the initial time step the Player selects a point close to the vertical mid-point of $\nsp{0}{1}$ to balance the two possible convex regions $\nsp{1}{1}$ and $\nsp{1}{3}$ at the next time step.
%
%
The Opponent is then indifferent regarding its next state selection between nodes 1 and 3.
Similarly, at time step $1$, the Player selects a horizontal mid-point in $\nsp{1}{1}$ to balance between the convex regions $\nsp{2}{1}$ and $\nsp{2}{3}$ that may be selected at time step 2.
Finally, \figref{fig:unilateral-deviation} depicts the scenario when one of the agents slightly deviates from its discretized optimal policy, leading to a suboptimal $\widehat{\opponentV}_0$.
One can see an increase in the cost when the Player deviates and a decrease in the cost when the Opponent deviates. 
Notice that when the Player deviates and does not select a vertical mid-point in $\nsp{1}{1}$, the Opponent counteracts and selects $\nsp{3}{1}$ to maximize the Player's error.

\section{Conclusion} \label{sec:conclusion}

In this work, we have extended the convex body chasing problem to an adversarial setting, where a Player chases a sequence of convex bodies assigned adversarially by an Opponent. 
We showed that under the assumption that the set of convex bodies is finite and known to both agents, max-min optimal policies can be obtained, which have a stronger performance guarantee than those in the classical CBC literature that use the competitive ratio. 
We showed how to compute the optimal value of the game and proved its continuity under certain assumptions. 
A discretization scheme has also been proposed to  numerically solve for the value function of the game with
performance guarantees for the corresponding discretized policies.
Numerical examples verified the theoretical developments. 
Future work will address the case of the probabilistic selection of the Opponent of the convex sets, and the extension of the numerical solution beyond box convex sets to address more realistic scenarios.
We also plan to
utilize this framework to introduce a cost structure to adversarial resource allocation problems, such as the dynamic Defender-Attacker Blotto Games~\cite{shishika2021dynamic}.


\clearpage
\bibliographystyle{ieeetr}
\bibliography{refs}


\appendix

\clearpage
\begin{appendices}
    \section{Proof of Lemma~\ref{lmm:intersection-continuity}}
    \label{appdx:intersect_correspondence_cont}
    \begin{proof}
        From Defintion~\ref{def:set-valued map continuity}, it follows that $\Gamma$ is upper semi-continuous (usc).
Let the set-valued map $\widetilde{F} : C \rightsquigarrow \Y$ such that $\widetilde{F}(\x) = F$ for all $\x \in C$.
 Notice that 
 $\mathrm{Graph}(\widetilde{F}) = \{ (\x,\z) \in C \times \Y : \z \in \widetilde{F}(\x) = F\} = C \times F$ is closed in $\X \times \Y$.
    %
        By Corollary~2.12 in \cite{freeman2008robust}, the correspondence $\Gamma$ being usc with compact values on $C$ and $\mathrm{Graph}(\widetilde{F})$ being closed imply that $\Xi: C \rightsquigarrow \Y$ is usc. 
    
        It remains to show that $\Xi$ is also lower semi-continuous (lsc) on $C$.
        Fix $\x \in C$, and consider an open set $E \subseteq \Y$ such that $E \cap \Xi(\x) \neq \emptyset$.
        Since $F$ and $\Gamma(\x)$ are convex and closed, the set $\Xi(\x)$ is also closed and convex.
        By the assumption that $\mathrm{int}(\Xi(\x)) \neq \emptyset$, $E \cap \Xi(\x) \neq \emptyset$ implies the open set $E \cap \mathrm{int}(\Xi(\x))$ is also nonempty. 
        %
    %
        Let now $G$ be a non-empty open set such that 
        \begin{equation}
            G \subseteq E \cap \mathrm{int}(\Xi(\x)) = E \cap \mathrm{int} (\Gamma(\x) \cap F) \subseteq E \cap ( \Gamma(\x) \cap F ) = E \cap \Xi(\x).
            \label{eqn:open set in E and int(Xi)}
        \end{equation}

        From (\ref{eqn:open set in E and int(Xi)}), it follows that the non-empty open set $G$ is a subset of $\Gamma(\x)$, hence $\Gamma(\x) \cap G \neq \emptyset$.
        %
        %
        By lower semi-continuity of $\Gamma$ on $C$, there exists a neighborhood $H$ of $\x$ such that $\Gamma(\x') \cap G \neq \emptyset$ for all $\x' \in H$. 
        Since $G \subseteq F$ from (\ref{eqn:open set in E and int(Xi)}), we also have
        \begin{equation}
            \Gamma(\x') \cap G = G \cap \Gamma(\x') \cap F = G \cap \Xi(\x') \neq \emptyset.
        \end{equation}
        Furthermore, since $G \subseteq E$ as shown in (\ref{eqn:open set in E and int(Xi)}), it follows that $E \cap \Xi(\x') \neq \emptyset$ for all $\x' \in H$.
        Therefore, for any open set $E \subseteq \Y$  such that $\Xi(\x) \cap E \neq \emptyset$, there exists a neighborhood $H$ of $\x$ such that $\Xi(\x') \cap E \neq \emptyset$ for all $\x' \in H$. 
        By Definition~\ref{def:set-valued map lsc}, $\Xi$ is lsc on $C$.
    
        Finally, upper and lower semi-continuity of correspondence $\Xi$ on $C$ imply the continuity of $\Xi$ on $C$ by Definition~\ref{def:set-valued map continuity}.
        %
    \end{proof}

    
    \section{Proof of Theorem~\ref{thm:value_fx_cont}}
    \label{appdx:lemma_value_uniform_cont}   
    
    \begin{proof}
    We only need to show the continuity of the value functions $\playerV_t$ and $\opponentV_t$ with $t \geq 1$.
    %
    We first prove the 
    continuity of the Player's value function $\playerV_t$ by induction. 
    
    
    \textit{Base case:} Consider $t = T$, fix $i \in \nodeset$ and let $j \in \neighbor{i}$. 
    %
    %
    The continuity of cost function $c: \X \times \X \rightarrow \R$ and the compactness of
    $\icj{T}{j}$
    imply that the infimum in (\ref{eqn:V_T_def}) is attainable and finite for any fixed $\x_{T-1} \in \nsp{T-1}{i}$. 
    Hence, \eqref{eqn:V_T_def} can be written as
    \begin{equation*}
        \playerV_{T}~(\x_{T-1}, j) = \min_{\x_{T} ~ \in ~ \icj{T}{j}}{ c~(\x_{T-1}, \x_{T})}.
    \end{equation*}
    From Lemma~\ref{lmm:Theta Cont}, we know that the compact-valued
    %
    %
    correspondence $\Theta_T^{(j)}$ is continuous on $\nsp{T-1}{i}$.
    Together with the continuity of the cost function $c$ on $\nsp{T-1}{i} \times \X$,  Lemma~\ref{lmm:marginal-fx-cont} implies that the Player's value function $\playerV_T~(\cdot, j): \nsp{T-1}{i} \to \R$ is continuous. 
    %
    %
    %
    %
    
    
    \textit{Inductive hypothesis:} Let some $t\in\{1, \ldots, T-1\}$, and suppose that $\playerV_{t+1}~(\cdot, k): \nsp{t}{j} \to \R$ 
    is continuous 
    for all $j \in \nodeset$ and $k \in \neighbor{j}$.
    
    \textit{Induction step:} Fix $i \in \nodeset$ and let $j \in \neighbor{i}$.
    We want to show that $\playerV_{t}~(\cdot, j): \nsp{t-1}{i} \to \R$ is continuous. 
    %
    Since $\playerV_{t+1}~(\cdot, k)$ is always continuous on $\nsp{t}{j}$ for all $k \in \neighbor{j}$, Lemma~\ref{lmm: max_fx_uniform_cont} implies that the function $g_t(\cdot, j): \nsp{t}{j} \to \R$ characterized by
    \begin{equation*}
        g_t(\x_t, j) \triangleq \max_{k ~ \in ~ \neighbor{j}} ~ \playerV_{t+1} ~(\x_t, k)
    \end{equation*}
    in \eqref{eqn:playerV_t} is also continuous. 
    %
    %
    Together with the continuity of the cost function $c$ on $\nsp{t-1}{i} \times \nsp{t}{j}$, one can further conclude that the function $f_t: \nsp{t-1}{i} \times \nsp{t}{j} \to \R$ characterized by
    %
    %
    %
    %
    \begin{equation*}
        f_t~(\x_{t-1}, \x_t) \triangleq c~(\x_{t-1}, \x_t) + g_t(\x_t, j)
    \end{equation*}
    is also continuous.
    %
    %
    %
    %
    Since $\icj{t}{j}$ is compact, the infimum in $\playerV_{t}~(\x_{t-1}, j)$ is attainable and finite for all $\x_{t-1} \in \nsp{t-1}{i}$. 
    Therefore, \eqref{eqn:V_t_def} can be written as 
    \begin{equation*}
        \playerV_t~(\x_{t-1}, j) = \min_{\x_t \in \icj{t}{j}}{\left\{c~(\x_{t-1}, \x_t) + \max_{k~\in~\neighbor{j}}\playerV_{t+1}~(\x_t,k)\right\}} = \min_{\x_t \in \icj{t}{j}}{f_t~(\x_{t-1}, \x_t)}.
    \end{equation*}

    From the continuity of $f_t$ and the compact-valued correspondence $\Theta_{t}^{(j)}$ on $\nsp{t-1}{i} \times \nsp{t}{j}$ and $\nsp{t-1}{i}$ respectively, Lemma~\ref{lmm:marginal-fx-cont} implies that the Player's value function $\playerV_t(\cdot, j)$ is continuous on $\nsp{t-1}{i}$.
    This completes the induction step.
    %
    %
    %
    %
    %
    Based on the relation  between $\playerV_t$ and $\opponentV_t$ in Lemma~\ref{lmm:U=max_V},
    for all $t \in \{1, \ldots, T\}$ and $i \in \nodeset$,
    the continuity of $\playerV_t(\cdot, j)$ on $\nsp{t-1}{i}$ for all $j \in \neighbor{i}$ implies the continuity of $\opponentV_t(\cdot, i): \nsp{t-1}{i} \to \R$ as a direct consequence of Lemma~\ref{lmm: max_fx_uniform_cont}.

\end{proof}

\subsection{Supporting Results for Theorem~\ref{thm:value_fx_cont}}
Recall the following definition of the Player's value function $\playerV_t$. 
    
\begin{equation*}
    \playerV_t~(\x_{t-1}, i_t) = \inf_{\x_t ~\in ~\ic{t}} \left\{c~(\x_{t-1}, \x_t) + \max_{i_{t+1} ~\in~ \neighbor{i_t}} \playerV_{t+1}~(\x_t, i_{t+1})\right\}.
\end{equation*}
Notice that the optimization domain depends on $\x_{t-1}$, which is characterized by the  correspondence $\Theta_t^{(i_t)}$. 
In order to show that $V_t$ is continuous with respect to $\x_{t-1}$, we need to first ensure the continuity of the correspondence $\Theta_t^{(i_t)}$.

The following lemma provides us with the desired continuity property of $\Theta_t^{(i_t)}$.

\begin{lemma}
    For all $t \in \{1,\ldots, T\}$ and $i_t \in \nodeset$, 
    the correspondence $\Theta_t^{(i_t)}$
    is continuous on $\nsp{t-1}{j}$ for all $j$ such that $i_{t} \in \neighbor{j}$.
    \label{lmm:Theta Cont}
\end{lemma}

\begin{proof}
    Fix $t \in \{1, \ldots, T\}$ and $i_t \in \nodeset$. 
    Under Assumptions~\ref{assmp:R set compact} and \ref{assmp:R set Hausdorff Cont}, we know that the reachability correspondence $\RSet: \X \rightsquigarrow \X$ is continuous, and $\RSet(\x)$ is compact and convex.
    Under Assumptions~\ref{assmp:convex body} and \ref{assmp:feasibility}, for the closed and convex set $\nsp{t}{i_{t}}$, we have $\mathrm{int}(\RSet(\x_{t-1}) \cap \nsp{t}{i_t}) \neq \emptyset$ for all $\x_{t-1} \in \nsp{t-1}{j}$ where $i_t \in \neighbor{j}$.
    The continuity of $\Theta_t^{(i_t)}$ then follows directly from Lemma~\ref{lmm:intersection-continuity}.
\end{proof}
    
\begin{lemma}
Let $\X \subseteq \R^d$ and $N \geqslant 2$, and let, for all $i\in \{1, \ldots, N\}$,
$f_i:\X \to \R$ be continuous.  
Then, the function ${f_{\max}(\x) = \max_{i\in \{1, \ldots, N\}} f_i (\x)}$ is continuous.
\label{lmm: max_fx_uniform_cont}
\end{lemma}

\begin{proof}
    The lemma can be easily proved using the identity $\max \{a, b\} = \frac{1}{2}\big((a+b) - \abs{a-b}\big)$.
    %
\end{proof}

\section{Proof of Lemma~\ref{lmm:U=max_V}}
\label{appdx:UV-relation}
\UV*

\begin{proof}
    We first apply induction to prove that \eqref{eqn:Ut=max Vt} holds for all $t \in \{1, \ldots, T\}$ and $i_{t-1} \in \nodeset$.
    
    \textit{Base Case:} 
    At $t = T$, (\ref{eqn:playerV_T}) and (\ref{eqn:opponentV_T}) directly imply that for all $i_{T-1} \in \nodeset$,
    \begin{equation*}
        \opponentV_T~(\x_{T-1}, i_{T-1}) = \max_{i_T ~\in~ \neighbor{i_{T-1}}} \playerV_T~(\x_{T-1}, i_T).
    \end{equation*}
    
    \textit{Inductive Hypothesis:}
    Suppose at some $t \in \{2, \ldots, T\}$, \eqref{eqn:Ut=max Vt} holds for all ${i_{t-1} \in \nodeset}$.
    
    \textit{Induction Step:} We want to show that \eqref{eqn:Ut=max Vt} also holds for all $i_{t-2} \in \nodeset$.
    It follows from \eqref{eqn:opponentV_t} that, for all $i_{t-2} \in \nodeset$, 
    \begin{equation*}
        \opponentV_{t-1}~(\x_{t-2}, i_{t-2}) = \max_{i_{t-1} ~\in ~\neighbor{i_{t-2}}} \left\{\min_{\x_{t-1} ~\in ~ \Theta_{t-1}^{(i_{t-1})}(\x_{t-2})} \{c~(\x_{t-2}, \x_{t-1}) + \opponentV_{t}~(\x_{t-1}, i_{t-1})\}\right\}.
    \end{equation*}
    
    Replacing $\opponentV_{t}~(\x_{t-1}, i_{t-1})$ in the above equation with the assumed relation in the inductive hypothesis and combining with \eqref{eqn:playerV_t}, yields
    \begin{align*}
        \opponentV_{t-1}~(\x_{t-2}, i_{t-2}) &= \max_{i_{t-1} ~\in ~\neighbor{i_{t-2}}} \left\{\min_{\x_{t-1} ~\in ~ \Theta_{t-1}^{(i_{t-1})}(\x_{t-2})} \{c~(\x_{t-2}, \x_{t-1}) + \max_{i_t ~\in ~\neighbor{i_{t-1}}} \playerV_t~(\x_{t-1}, i_t)\}\right\}\\
        &= \max_{i_{t-1} ~\in ~\neighbor{i_{t-2}}}\playerV_{t-1} ~(\x_{t-2}, i_{t-1}).
    \end{align*}

    We conclude that \eqref{eqn:Ut=max Vt} holds for all $i_{t-2} \in \nodeset$, which completes the induction step.
    
    %
    Furthermore, we have that $\opponentV_1~(\x_0, i_0) = \max_{i_1 ~\in~ \neighbor{i_0}} \playerV_1~(\x_0, i_1)$ for all $i_0 \in \nodeset$.
    By expressing $\opponentV_1$ in terms of $\playerV_1$, \eqref{eqn:playerV_0} and \eqref{eqn:opponentV_0} directly indicate $\opponentV_0~(\nodeset, \Q) = \max_{i_0 ~\in~ \nodeset} \playerV_0~(i_0)$.
    After proving the relation in \eqref{eqn:Ut=max Vt} and \eqref{eqn:U0 in terms of V0}, the Opponent value's relation to the Player value represented by \eqref{eqn:Vt = min (c + Ut+1)} and \eqref{eqn:V0 in terms of U1} are direct consequences of substituting \eqref{eqn:Ut=max Vt} into \eqref{eqn:playerV_t} and \eqref{eqn:playerV_0} respectively.
    %
\end{proof}

\section{Proof of Theorem~\ref{thm:value_fx_L_cont}}
\label{appdx:lemma_value_L_cont}
\begin{proof}
    We will only prove the above result for $V_t$ through induction since the case for $U_t$ can be easily obtained from the relations between $\playerV_t$ and $\opponentV_t$ using Lemma~\ref{lmm: max_fx_lip_cont}. 
    
    \textit{Base case: } Let $t=T$, 
    and let $i_{T-1} \in \nodeset$ and $i_T \in \neighbor{i_{T-1}}$.
    From Assumption~\ref{assmp:feasibility}, we know that $\icjx{T}{i_T}{\x} \neq \emptyset$ for all $\x \in \nsp{T-1}{i_{T-1}}$.
    Then, for all $\x_{T-1}$, $\x'_{T-1} \in \nsp{T-1}{i_{T-1}}$,we have
    \begin{align*}
        \abs{V_T(\x_{T-1},i_T) - V_T(\x'_{T-1},i_T)} &= \abs{\min_{\x_T \in \ic{T}} c(\x_{T-1},\x_T) - \min_{\x'_T \in \Theta_T^{(i_T)}(\x'_{T-1})} c(\x'_{T-1},\x'_T)}\\
        &\leq \underbrace{\abs{\min_{\x_T \in \ic{T}} c(\x_{T-1},\x_T) - \min_{\x_T \in \ic{T}} c(\x'_{T-1},\x_T)}}_{A}\\
       & + \underbrace{\abs{\min_{\x_T \in \ic{T}} c(\x'_{T-1},\x_T) - \min_{\x'_T \in \Theta_T^{(i_T)}(\x'_{T-1})} c(\x'_{T-1},\x'_T)}}_{B}.
    \end{align*}
    By Assumption~\ref{assmp:cost-L-cont}, we have $\abs{c(\x_{T-1}, \x_{T}) - c(\x'_{T-1}, \x_{T})} \leq L_c \norm{\x_{T-1}-\x'_{T-1}}$ for all $\x_T \in \ic{T}$. 
    Consequently, and since $\ic{T}$ is compact,
    Lemma~\ref{lmm:min-cont} implies that $A \leq L_c \norm{\x_{T-1}-\x'_{T-1}}$.
    Since $c~(\x_{T-1}', \cdot)$ is Lipschitz continuous with respect to $\x_T \in \nsp{T}{i_T}$ and the compact-valued correspondence $\Theta_T^{(i_T)}$ is Lipschitz under the Hausdorff distance by Assumption~\ref{assmp:R-set-L-cont},     Lemma~\ref{lmm:min_hausdorff} implies that $B \leq L_\Theta L_c \norm{\x_{T-1}-\x'_{T-1}}$.
    Consequently, we have
    \begin{equation*}
        \abs{V_T(\x_{T-1},i_T) - V_T(\x'_{T-1},i_T)} \leq L_c (1+ L_\Theta) \norm{\x_{T-1}-\x'_{T-1}} = L_{v, T} \norm{\x_{T-1} - \x'_{T-1}}.
    \end{equation*}
    
    \textit{Inductive Hypothesis:}
    Suppose at some $t \in \{1, \ldots, T-1\}$, $\playerV_{t+1}~(\cdot, i_{t+1})$ is $L_{v, t+1}$-Lipschitz continuous on $\nsp{t}{i_t}$ for all $i_t \in \nodeset$ and $i_{t+1} \in \neighbor{i_t}$.
    
    \textit{Induction Step:}
    Fix $i_{t-1} \in \nodeset$ and let $i_t \in \neighbor{i_{t-1}}$.
    Recalling \eqref{eqn:playerV_t}, we have
    \begin{equation*}
        \playerV_{t}~(\x_{t-1}, i_{t}) = \min_{\x_{t} \in \ic{t}} \left\{c~(\x_{t-1}, \x_{t}) + \max_{i_{t+1} \in \neighbor{i_t}} \playerV_{t+1}(\x_{t}, i_{t+1})\right\}.
    \end{equation*}
    From the inductive hypothesis and Lemma~\ref{lmm: max_fx_lip_cont}, we have that $U_{t+1}(\x_{t}, i_{t}) = \max_{i_{t+1} \in \neighbor{i_t}} \{V_{t+1}(\x_{t}, i_{t+1})\}$ is also $L_{v,t+1}$-Lipschitz with respect to $\x_t \in \nsp{t}{i_t}$.
    By repeating the process as in the base case using Lemma~\ref{lmm:min-cont} and Lemma~\ref{lmm:min_hausdorff}, and
    combining with the fact that $c$ is $L_c$-Lipschitz continuous with respect to $\x_t$
   yields that $V_{t}(\x_{t-1}, i_{t})$ is Lipschitz continuous with Lipschitz constant $(L_c +L_{v,t+1}) (1+L_\Theta)$. 
    Plugging in the expression of $L_{v,t+1}$, finally yields
    \begin{equation*}
        L_{v, t} = (L_c +L_{v,t+1}) (1+L_\Theta) = (L_c + L_c \sum_{k=1}^{T-t} (1+L_\Theta)^k)(1+L_\Theta) = L_c \sum_{k=1}^{T-t+1} (1+L_\Theta)^k,
    \end{equation*}
    which completes the induction.
\end{proof}

\subsection{Supporting Results for Theorem~\ref{thm:value_fx_L_cont}}

\begin{lemma}
\label{lmm:min-cont}
Let a compact set $\X \subseteq \R^d$ and continuous functions $f, g : \X \to \R$. 
Let $\epsilon \geq 0$, and suppose that, for all $\x\in \X$, $|f(\x) - g(\x)| \leq \epsilon$.
Then, $|\min_{\x\in \X}f(\x) - \min_{\x \in \X} g(\x)| \leq \epsilon$.
\end{lemma}

\begin{proof}
Let $\widehat{\x} \in \argmin_{\x\in \X}{f(\x)}$.
It follows that
\begin{equation*}
    \min_{\x \in \X} g(\x) - \epsilon \leq g(\widehat{\x}) - \epsilon \leq f(\widehat{\x}) = \min_{\x \in \X} f(x). 
\end{equation*}
Similarly, let $\widetilde{\x} \in \argmin_{\x\in\X}{g(\x)}$, so that
\begin{equation*}
    \min_{\x \in \X} f(\x) \leq f(\widetilde{\x}) \leq g(\widetilde{\x}) +\epsilon = \min_{\x \in \X} g(\x) +\epsilon. 
\end{equation*}
In follows immediately that
\begin{equation*}
    \Big|\min_{\x\in \X}f(\x) - \min_{\x\in \X} g(\x)\Big| \leq \epsilon.
\end{equation*}
\end{proof}

\begin{lemma}
Let $\X \subseteq \R^d$ and $N \geqslant 2$, and let $f_i:\X \to \R$ be $L_i$-Lipschitz continuous for all $i\in \{1, \ldots, N\}$.
Then, the function $f_{\max}(\x) = \max_{i\in\{1,\ldots, N\}} f_i(\x)$ is $L$-Lipschitz continuous with $L = \max_{i} L_i$.
\label{lmm: max_fx_lip_cont}

\end{lemma}

\begin{proof}
    We will only prove the case where $N=2$. 
    The case $N>2$ can be shown by induction.
    Let $\x, \x' \in \X$.
    By the Lipschitz continuity of $f_1$ and $f_2$ we have that
    \begin{align*}
        f_1(\x') &\leq f_1(\x) + L_1 \norm{\x-\x'} \leq f_{\max}(\x) + L  \norm{\x-\x'}, \\
        f_2(\x') &\leq f_2(\x) + L_2 \norm{\x-\x'} \leq f_{\max}(\x) + L \norm{\x-\x'}.
    \end{align*}
It follows that
    \begin{equation*}
        f_{\max}(\x') \leq f_{\max}(\x) + L  \norm{\x-\x'}.
    \end{equation*}
    By symmetry, we also have $f_{\max}(\x) \leq f_{\max}(\x') + L \norm{\x-\x'}$, and consequently the function $f_{\max}$ is $L$-Lipschitz continuous.
\end{proof}

\begin{lemma}
    \label{lmm:min_hausdorff}
    Consider a Lipschitz continuous function $f:\Y \to \mathbb{R}$ with Lipschitz constant $L_f$ and a compact-valued correspondence $\Gamma: \X \rightsquigarrow \Y$, which is Lipschitz continuous under the Hausdorff distance with Lipschitz constant $L_\Gamma$.
    Then, the real-valued function $\psi(\x) = \min_{\y\in \Gamma(\x)} f(\y)$ is also Lipschitz continuous with Lipschitz constant of $L_\psi = L_f  L_\Gamma$.
\end{lemma}

\begin{proof}
   Let $\x, \x' \in \X$. It follows that
    \begin{equation*}
        \abs{\psi(\x) - \psi(\x')} = \abs{\min_{\y \in \Gamma(\x)}f(\y) - \min_{\y \in \Gamma(\x')}f(\y)}.
    \end{equation*}

    The continuity of $f$ and the compactness of $\Gamma(\x)$ imply that the set of minima of $f$ over the domain $\left(\Gamma(\x) \cup \Gamma(\x')\right)$ is non-empty.
    Consequently, the minimum can be attained in either $\Gamma(\x)$, $\Gamma(\x')$, or both.
    Without loss of generality, consider the case where the minimum is attained in $\Gamma(\x)$. Formally,
    \begin{equation*}
        \left(\argmin_{\y \in \Gamma(\x) \cup \Gamma(\x')} f(\y)\right) \cap \Gamma(\x) \neq \emptyset.
    \end{equation*}
    
    In this case, there exists $\y^* \in \Gamma(\x)$ such that
    \begin{equation*}
        f(\y^*) = \min_{\y \in \Gamma(\x)} f(\y) = \min_{\y \in \Gamma(\x) \cup \Gamma(\x')} f(\y) \leq \min_{\y \in \Gamma(\x')} f(\y),
    \end{equation*}
    which implies that
    \begin{equation*}
        \abs{\psi(\x) - \psi(\x')} = \min_{\y \in \Gamma(\x')}f(\y) - \min_{\y \in \Gamma(\x)}f(\y) = \min_{\y \in \Gamma(\x')}f(\y)  - f(\y^*).
    \end{equation*}
    From the Lipschitz continuity of the correspondence $\Gamma$ and the definition of the Hausdorff distance,
    we have 
    \begin{equation*}
        \inf_{\y \in \Gamma(\x')} \norm{\y^* - \y} \leq 
        \mathrm{dist}_H(\Gamma(\x'), \Gamma(\x)) \leq L_\Gamma  \norm{\x-\x'}.
    \end{equation*}
    By the compactness of $\Gamma(\x')$,
   there exists
   $\widetilde{\y} \in \Gamma(\x')$
    such that 
    \begin{equation*}
        \norm{\y^* - \widetilde{\y}} = \inf_{\y \in \Gamma(\x')} \norm{\y^* - \y} \leq L_\Gamma \norm{\x - \x'}.
    \end{equation*}
    Together with the Lipschitz continuity of the function $f$, we conclude that, for all $\x$, $\x' \in \X$,
    \begin{equation*}
        \abs{\psi(\x) - \psi(\x')}= \min_{\y \in \Gamma(\x')}f(\y)  - f(\y^*) \leq f(\widetilde{\y}) - f(\y*) \leq L_f \norm{\widetilde{\y} - \y^*} \leq L_\Gamma L_f \norm{\x -\x'}.
    \end{equation*}
\end{proof}


\section{Theoretical Results on Discretization}

\subsection{Discretized Value Functions}   \label{appdx:discretized-value}

The following are the propagation rules for the discretized Player value functions. 
Notice that the $\x$-argument domain and the $\x$-optimization domain of the discretized value functions are characterized by the discretized state spaces $\{\widehatX_t\}_{t=0}^{T}$.
%
\begin{subequations}
    \label{eqn:apprx-player-V}
    \begin{align}
        \widehat{\playerV}_T~(\widehatx_{T-1}, i_T) &= \min_{\widehatx_{T} \, \in \,\widehatX_T \, \cap \, \widehatic{T}
        } c\;(\widehatx_{T-1}, \widehatx_T),
        \label{eqn:apprx-playerV_T}
        \\
         \widehat{\playerV}_t~(\widehatx_{t-1}, i_t) &= \min_{\widehatx_t \, \in \, \widehatX_t \, \cap \, \widehatic{t}
         } \left\{c~(\widehatx_{t-1}, \widehatx_t) + \max_{i_{t+1} \, \in \, \neighbor{i_t}} \widehat{\playerV}_{t+1}~(\widehatx_t, i_{t+1})\right\},
         ~~\forall~t = 1,\ldots,T-1,
        \label{eqn:apprx-playerV_t}
        \\
        \widehat{\playerV}_0~(i_0) &= \min_{\widehatx_0 ~\in~ \widehatX_0 \, \cap \, \nsp{0}{i_0}} ~ \max_{i_1 ~\in ~\neighbor{i_0}} \widehat{\playerV}_1~(\widehatx_0, i_1).
     \end{align}
\end{subequations}
Similarly, 
\begin{subequations}
    \label{eqn:apprx-opponent-V}
    \begin{align}
        \widehat{\opponentV}_T~(\widehatx_{T-1}, i_{T-1}) &= \max_{i_T \, \in \, \neighbor{i_{T-1}}} \left\{\min_{\widehatx_T \, \in \, \widehatX_T \, \cap \, \widehatic{T}
        } c~(\widehatx_{T-1}, \widehatx_T)\right\},
        \label{eqn:apprx-opponentV_T}\\
        \widehat{\opponentV}_t~(\widehatx_{t-1}, i_{t-1}) &= \max_{i_t \, \in \, \neighbor{i_{t-1}}} \left\{\min_{\widehatx_t \, \in \, \widehatX_{t} \, \cap \, \widehatic{t}
        } \{c~(\widehatx_{t-1}, \widehatx_t) + \widehat{\opponentV}_{t+1}~(\widehatx_t, i_t)\}\right\},
         ~~ \forall~t =1,\ldots, T-1,
        \label{eqn:apprx-opponentV_t} \\
        \widehat{\opponentV}_0~(\nodeset, \Q) &= \max_{i_0 \, \in \, \nodeset} ~\min_{\widehatx_0 \, \in \, \widehatX_{0} \, \cap \, \nsp{0}{i_0}} \widehat{\opponentV}_1~(\widehatx_0, i_0).
        \label{eqn:apprx-opponentV_0}
    \end{align}
\end{subequations}

\subsection{Discretized Policies}   \label{appdx:discretized-policy}

The discretized optimal Player policies are defined as
\begin{subequations}
\label{eqn:apprx-player-optimal-policy}
    \begin{align}
        \widehat{\pi}_T^*~(\widehatx_{T-1}, i_T) &\in \argmin_{\widehatx_T \,\in \, \widehatX_T \, \cap \, \widehatic{T} 
        } c~(\widehatx_{T-1}, \widehatx_{T}), \label{eqn:apprx-player-optimal-policy-T}\\
        \widehat{\pi}_t^*~(\widehatx_{t-1}, i_t) &\in \argmin_{\widehatx_t \, \in \, \widehatX_t \, \cap \, \widehatic{t}
        } \{c~(\widehatx_{t-1}, \widehatx_t) + \widehat{\opponentV}_{t+1}~(\widehatx_t, i_t)\}, \quad \forall \; t=1,\ldots,T-1, \label{eqn:apprx-player-optimal-policy-t}\\
        \widehat{\pi}_0^*~(i_0) &\in \argmin_{\widehatx_0 \, \in \, \widehatX_0 \, \cap \, \nsp{0}{i_0}} \widehat{\opponentV}_1~(\widehatx_0, i_0). \label{eqn:apprx-player-optimal-policy-0}
    \end{align}
\end{subequations}

Similarly, the optimal discretized Opponent policy is defined as
\begin{subequations}
\label{eqn:apprx-opponent-optimal-policy}
    \begin{align}
    \widehat{\sigma}_T^*~(\widehatx_{t-1},i_{t-1}) &\in \argmax_{i_t \, \in \, \neighbor{i_{t-1}}} \widehat{\playerV}_t~(\widehatx_{t-1}, i_t), \quad \forall \; t=1,\ldots,T, \label{eqn:apprx-opponent-optimal-policy-t}\\
    \widehat{\sigma}_0^*~(\mathcal{G}, \Q) &\in \argmax_{i_0 \, \in \, \nodeset} \widehat{\playerV}_0 ~ (i_0). \label{eqn:apprx-opponent-optimal-policy-0}
\end{align}
\end{subequations}

\subsection{Discretization Error Bounds}    \label{appdx:approximation-theorem}

\appxv*

\begin{proof}
    We will prove this theorem by induction.
    
    \textit{Base case:} 
    Let $t = T$, let $i_{T-1} \in \nodeset$, and $\widehat{\x}_{T-1} \in \widehat{\X}_{T-1} \cap \nsp{T-1}{i_{T-1}}$, and let $i_T \in \neighbor{i_{T-1}}$.
    From \eqref{eqn:playerV_T} and \eqref{eqn:apprx-playerV_T}, we have that
    \begin{equation*}
        \playerV_T~(\widehat{\x}_{T-1}, i_T) \leq \widehat{\playerV}_T~(\widehat{\x}_{T-1}, i_t).
    \end{equation*}

    Let $\x_T^* \! \in \! \argmin_{\x_T \in \widehatic{T}} c\,(\widehatx_{T-1}, \x_T)$ be the optimal Player action at time $T$.
    It follows from \eqref{eqn:discretization-intersection} and the Lipschitz continuity of the cost function $c$ that there exists $\widehat{\x}_T^* \in \widehatX_T \cap \widehatic{T}$ such that 
    \begin{equation*}
        \abs{c~(\widehat{\x}_{T-1}, \x_T^*) - c~(\widehat{\x}_{T-1}, \widehat{\x}_T^*)} \leq L_c \delta_{\X, T}.
    \end{equation*}
    Consequently, using \eqref{eqn:playerV_T} and \eqref{eqn:apprx-playerV_T}, we have
    \begin{equation*}
        \widehat{\playerV}_T~(\widehatx_{T-1}, i_T) \leq \playerV_T~(\widehatx_{T-1}, i_T) + L_c \delta_{\X, T}.
    \end{equation*}
    Using Lemma~\ref{lmm:U=max_V},
    it can be easily obtained that
    \begin{equation*}
        \opponentV_T~(\widehatx_{T-1}, i_{T-1}) \leq \widehat{\opponentV}_T~(\widehatx_{T-1}, i_{T-1}) \leq \opponentV_T~(\widehatx_{T-1}, i_{T-1}) + L_c \delta_{\X, T}.
    \end{equation*}
    
    \inductivehypothesis
    Suppose at some time step $t \in \{2, \ldots, T\}$, \eqref{eqn:Ut and hatUt inequality} and \eqref{eqn:Vt and hatVt inequality} hold for all $\widehatx_{t-1} \in \widehatX_{t-1} \cap \nsp{t-1}{i_{t-1}}$.
    
    \induction
    For ease of notation, we first define 
    \begin{equation*}
        \widehat{\varepsilon}_{t} = L_c \delta_{\X, T} + \sum_{\tau= t}^{T-1} (L_c+L_{v,\tau+1})  \delta_{\X,\tau}.
    \end{equation*}
    
    Now fix $i_{t-2} \in \nodeset$ and $\widehatx_{t-2} \in \widehatX_{t-2} \cap \nsp{t-2}{i_{t-2}}$,
    and let $i_{t-1} \in \neighbor{i_{t-2}}$.
    From the inductive hypothesis and \eqref{eqn:apprx-playerV_t}, it follows that 
    \begin{equation*}
        \playerV_{t-1}~(\widehatx_{t-2}, i_{t-1}) \leq \widehat{\playerV}_{t-1}~(\widehatx_{t-2}, i_{t-1}).
    \end{equation*}
    
    Moreover, notice that
    \begin{align*}
        \widehat{\playerV}_{t-1}~(\widehat{\x}_{t-2}, i_{t-1}) & \leq \min_{\widehat{\x}_{t-1} \, \in \, \widehat{\X}_{t-1} \, \cap \, \Theta_{t-1}^{(i_{t-1})}(\widehatx_{t-2})
        } \left\{c~(\widehat{\x}_{t-2}, \widehat{\x}_{t-1}) + \opponentV_t~(\widehat{\x}_{t-1}, i_{t-1})\right\} + \widehat{\varepsilon}_t\\
        &\leq \playerV_{t-1}~(\widehatx_{t-2}, i_{t-1}) + (L_c + L_{v,t}) \delta_{\X, t-1} + \widehat{\varepsilon}_t\\
        &= \playerV_{t-1}~(\widehat{\x}_{t-2}, i_{t-1}) + \widehat{\varepsilon}_{t-1},
    \end{align*}
    where the first inequality is the result of \eqref{eqn:apprx-playerV_t} from the inductive hypothesis on $\widehat{\opponentV}_t$.
    The second inequality is a consequence of \eqref{eqn:discretization-intersection} from the Lipschitz continuity of $\opponentV_t$ and $c$ with respect to 
    $\widehat{\x}_{t-1}$-argument.
    
    Using a similar argument  as in the base case, we then have
    \begin{equation*}
        \widehat{\opponentV}_{t-1}~(\widehatx_{t-2}, i_{t-2}) \leq \opponentV_{t-1}~(\widehatx_{t-2}, i_{t-2}) \leq \widehat{\opponentV}_{t-1}~(\widehatx_{t-2}, i_{t-2}) + \widehat{\varepsilon}_{t-1}.
    \end{equation*}
    This completes the induction.
\end{proof}

\appxvi*
\begin{proof}
    Fix $i_0 \in \nodeset$. 
    From Lemma~\ref{lmm:U=max_V}  
    we have
    \begin{align}
        \playerV_0~(i_0) &= \min_{\x_0 \, \in \, \nsp{0}{i_0}} \opponentV_1~(\x_0, i_0),\\
        \widehat{\playerV}_0~(i_0) &= \min_{\widehat{\x}_0 \, \in \, \widehat{\X}_0 \, \cap \, \nsp{0}{i_0}} \widehat{\opponentV}_1~(\widehatx_0, i_0).
    \end{align}
    Combining the above value function relations with \eqref{eqn:Ut and hatUt inequality} in Theorem~\ref{thm:discretization-error-bounds}, we have
    \begin{equation*}
        \playerV_0~(i_0) \leq \min_{\widehatx_0 \, \in \, \widehatX_0 \, \cap \, \nsp{0}{i_0}} \opponentV_1~(\widehatx_0, i_0) \leq \widehat{\playerV}_0~(i_0),
    \end{equation*}
    and 
    \begin{align}
        \widehat{\playerV}_0~(i_0) &\leq \min_{\widehatx_0 \, \in \, \widehatX_0 \, \cap \, \nsp{0}{i_0}} {\opponentV_1~(\widehatx_0, i_0)} + L_c \delta_{\X, T} + \sum_{\tau = 1}^{T-1} {(L_c + L_{v, \tau+1})\delta_{\X, \tau}} \label{eqn:1st ineq due to hatU1 and U1 inequality}\\
        &\leq \playerV_0~(i_0) + L_{v, 1} \delta_{\X, 0} + L_c \delta_{\X, T} + \sum_{\tau = 1}^{T-1} {(L_c + L_{v, \tau+1})\delta_{\X, \tau}} \label{eqn:2nd ineq due to Lipschitz in U1 wrt x0}.
    \end{align}
    Inequality \eqref{eqn:1st ineq due to hatU1 and U1 inequality} is the result of \eqref{eqn:Ut and hatUt inequality} in Theorem~\ref{thm:discretization-error-bounds}, while inequality \eqref{eqn:2nd ineq due to Lipschitz in U1 wrt x0} results from \eqref{eqn:discretization-intersection-0} and the Lipschitz continuity of $\opponentV_1$ with respect to the $\widehatx_0$-argument.
    Therefore, for all $i_0 \in \nodeset$, we have
    \begin{equation*}
        \playerV_0~(i_0) \leq \widehat{\playerV}_0~(i_0) \leq \playerV_0~(i_0) + L_{v,1} \delta_{\X,0} + L_c \delta_{\X, T} + \sum_{\tau = 1}^{T-1} {(L_c + L_{v, \tau+1})\delta_{\X, \tau}}.
    \end{equation*}
    Using the relation between $\widehat{\opponentV}_0$ and $\widehat{\playerV}_0$ 
    one easily arrives at
    \begin{equation*}
        \opponentV_0~(\nodeset, \Q) \leq \widehat{\opponentV}_0~(\nodeset, \Q) \leq \opponentV_0~(\nodeset, \Q) + L_{v, 1} \delta_{\X,0} + \sum_{\tau= 1}^{T-1} (L_c+L_{v,\tau+1}) \delta_{\X,\tau} + L_c \delta_{\X, T}.
    \end{equation*}
\end{proof}

\section{Proofs Related to the Numerical Simulations} \label{appdx:proofs for numerical simulation}

In the numerical simulation example, the state space $\X$ and the given convex sets $\Q$ are both assumed to be two-dimensional boxes of the form of $[a, b] \times [c, d]$ where $a < b$ and $c < d$. 
The reachability correspondence $\RSet$ on $\X$ is defined via the $\infty$-norm as in \eqref{eqn:R(x) in Numerical Simulation}.

\begin{lemma} \label{lmm:Euclidean cost Lipschitz}
   Let the cost be defined by $c(\x,\y) = \|\x - \y \|_2$.
   This cost is  1-Lipschitz continuous,
    that is, for all $(\x, \y), (\x', \y') \in \X \times \X$, 
    \begin{equation*}
        \abs{c~(\x, \y) - c~(\x', \y')} \leq (\norm{\x-\x'}_2 + \norm{\y - \y'}_2).
    \end{equation*}
\end{lemma}

\begin{proof}
    For all $(\x, \y), (\x', \y') \in \X \times \X$, using the Triangle Inequality, we have that
    \begin{align*}
        \abs{c~(\x, \y) - c~(\x', \y')} & = \abs{\norm{\x - \y}_2 - \norm{\x' - \y'}_2} \\ 
        & \leq \norm{\x-\y - \x' + \y'}_2\\
        & \leq \norm{\x - \x'}_2 + \norm{\y - \y'}_2.
    \end{align*}
\end{proof}

Lemma~\ref{lmm:min_P = min_P_cap_Q P Q boxes} and Lemma~\ref{lmm:Box A(x) 1-Lipschitz} below are needed to prove the Lipschitz continuity of the reachability correspondence $\RSet$ and the intersection correspondence $\Theta_{t}^{(i)}$.

\begin{lemma}
    \label{lmm:min_P = min_P_cap_Q P Q boxes}
    Let $P$ and $Q$ be boxes in $\R^d$ such that $P \cap Q \neq \emptyset$.
    Then, for all $\q \in Q$, 
    \begin{equation}
        \label{eqn:argmin_P subseteq P cap Q}
        \argmin_{\p \,\in \,P} {\twonorm{\p - \q}} \subseteq P \cap Q.
    \end{equation}
    Consequently,
    \begin{equation}
        \label{eqn:min_P = min_(P cap Q)}
        \min_{\p \,\in\, P} \norm{\p - \q}_2 = \min_{\p \,\in \,P \,\cap\, Q} \norm{\p - \q}_2.
    \end{equation}
\end{lemma}

\begin{proof}
    Since $P$ and $Q$ are boxes in $\R^d$, we can write
    \begin{align*}
        P  = \prod_{k=1}^{d} {\left[a_k, b_k\right]},\qquad
        Q  = \prod_{k=1}^{d} {\left[c_k, d_k\right]},
    \end{align*}
    for some $a_k < b_k$ and $c_k < d_k$.
    Since boxes and norms are compact and continuous, respectively, it follows that $\argmin_{\p \, \in\, P}\twonorm{\p - \q} \neq \emptyset$ for all ${\q \in Q}$. 
    Finding $\min_{\p \,\in \,P}\twonorm{\p - \q}$ is equivalent to finding $\min_{p_k \,\in \,[a_k, b_k]} \abs{p_k - q_k}$ for all $k \in \{1, \ldots, d\}$.
    Moreover, ${P~ \cap~ Q \neq \emptyset}$ implies that $[a_k, b_k] ~\cap~ [c_k, d_k] \neq \emptyset$ for all $k \in \{1, \ldots, d\}$.
    
    Now, fix $\q \in Q$. 
    Let $\widehat{\p} \in \argmin_{\p \, \in \, P}\twonorm{\p - \q}$ and
    let $\K = \{ k \in \{1, \ldots, d\} : q_k \notin [a_k, b_k]\}$.
    For all $k \in \K$, we have 
    \begin{equation*}
        \widehat{p}_k \in \argmin_{p_k \,\in\, [a_k, b_k]}\abs{p_k - q_k} \subseteq \{a_k, b_k\}.
    \end{equation*}
    Since $[a_k, b_k] ~ \cap ~[c_k, d_k] \neq \emptyset$, then at least one of $a_k$ and $b_k$ is within the interval $[c_k, d_k]$.
    We will next show that, for all $k \in \K$, $\widehat{p}_k \in [c_k, d_k]$.
    Without loss of generality, assume $b_k \in [c_k, d_k]$ and $a_k \notin [c_k, d_k]$.
    Since $a_k < b_k$, $b_k \in [c_k, d_k]$ and $a_k \notin [c_k, d_k]$, then $a_k < c_k \leq b_k < q_k \leq d_k$, implying that $q_k - a_k > q_k - b_k > 0$. 
    Therefore, $b_k \in \argmin_{p_k \, \in \, [a_k, b_k]}\abs{p_k - q_k}$, we have $\widehat{p}_k = b_k$.
    We conclude that, for all $k \in \K$, we have $\widehat{p}_k \in [a_k, b_k] ~\cap~[c_k, d_k]$.
    
    For all $k \in \{1, \ldots, d\}$ such that $q_k \in [a_k, b_k]$, one can easily observe that $\widehat{p}_k = q_k$. 
    Hence, $\widehat{p}_k ~\in ~[a_k, b_k] \cap [c_k, d_k]$.
    Finally, for fixed $\q \in Q$, let $\widehat{\p} \in \argmin_{\p\, \in\, P}\twonorm{\p - \q}$. 
    We then have
    \begin{equation*}
        \widehat{\p} \in \prod_{k=1}^{d} ~\left([a_k, b_k] ~\cap ~[c_k, d_k]\right) = \left(\prod_{k=1}^{d}~ [a_k, b_k]\right) ~\bigcap ~ \left(\prod_{k=1}^{d}~[c_k, d_k]\right) = P \cap Q.
    \end{equation*}
    This proves \eqref{eqn:argmin_P subseteq P cap Q}, which further implies \eqref{eqn:min_P = min_(P cap Q)}.
\end{proof}

\begin{lemma}
    \label{lmm:Box A(x) 1-Lipschitz}
    The correspondence $\A: \R^d \rightsquigarrow \R^d$  defined as
    \begin{equation}
        \label{eqn:Box A(x) definition}
        \A(\x) = \{\y \in \R^d:\norm{\y - \x}_\infty \leq \rho\},
    \end{equation}
    is 1-Lipschitz continuous under the Hausdorff distance.
\end{lemma}

\begin{proof}
    From \eqref{eqn:Box A(x) definition}, one can write $\A(\x)$ as 
    \begin{equation}
        \label{eqn:A(x) box expression}
        \A(\x) = \prod_{k=1}^d \left[x_k - \rho, x_k + \rho\right],
    \end{equation}
    which is a box in $\R^d$.
   Let $\x, \x' \in \R^d$, and let $\mathbf{h} = \x' - \x$.
   Notice that $\A(\x') = \A(\x) + \mathbf{h}$, which implies that there exists $\y' \in \A(\x')$ such that $\twonorm{\y - \y'}=\twonorm{\mathbf{h}}$ for all $\y \in \A(\x)$.
   Consequently, 
    \begin{equation*}
        \min_{\y' \in \A(\x')} \twonorm{\y - \y'} \leq \twonorm{\x - \x'},
    \end{equation*}
    which further implies that
    \begin{equation*}
        \max_{\y \in \A(\x)}~ \min_{\y' \in \A(\x')}~\twonorm{\y - \y'} \leq \twonorm{\x - \x'}.
    \end{equation*}
    In a similar way, one can also show that
    \begin{equation*}
        \max_{\y' \in \A(\x')}~\min_{\y \in \A(\x)}~\twonorm{\y - \y'} \leq \twonorm{\x - \x'}.
    \end{equation*}
    From Definition~\ref{def:Hausdorff distance def}, we have that, for all $\x, \x' \in \R^d$,
    \begin{equation*}
        \distH{\A(\x), \A(\x')} \leq \twonorm{\x - \x'}.
    \end{equation*}
    This completes the proof.  
\end{proof}

\begin{lemma}     \label{lmm:R(x) 1-Lipschitz Numerical Simulation}
    The reachability correspondence $\RSet: \X \rightsquigarrow \X$ in \eqref{eqn:R(x) in Numerical Simulation} is 1-Lipschitz continuous under the Hausdorff distance. 
    That is, for all $\x, \x' \in \X$,
    \begin{equation*}
        \distH{\RSet(\x), \RSet(\x')} \leq \twonorm{\x - \x'}.
    \end{equation*}
\end{lemma}

\begin{proof}
    From \eqref{eqn:R(x) in Numerical Simulation} and \eqref{eqn:Box A(x) definition} one can write $\RSet(\x)$ as follows:
    \begin{equation*}
        \RSet(\x) = \left(\prod_{k=1}^d \left[x_k - \rho, x_k + \rho\right]\right) \cap \X = \A(\x) \cap \X,
    \end{equation*}
    where $\A(\x)$ and $\X$ are both boxes in $\R^d$.
    For all $\x, \x' \in \X$, one obtains that $\x \in \RSet(\x)$ and $\x' \in \RSet(\x')$.
    Therefore, $\A(\x) \cap \X$ and $\A(\x') \cap \X$ are both non-empty.
    From Lemma~\ref{lmm:min_P = min_P_cap_Q P Q boxes}, and for all $\y' \in \X$, we have that
    \begin{equation*}
        \min_{\y \in \A(\x)} \twonorm{\y - \y'} = \min_{\y \in \RSet(\x)} \twonorm{\y - \y'}.
    \end{equation*}
    Combining the above results, along with Lemma~\ref{lmm:Box A(x) 1-Lipschitz}, we have, for all $\y' \in \RSet(\x')$, that
    \begin{align*}
        \min_{\y \in \RSet(\x)} \twonorm{\y - \y'} &\leq
        \max_{\y' \in \RSet(\x')}~\min_{\y \in \RSet(\x)}~ \twonorm{\y - \y'} \\
        & = \max_{\y' \in \RSet(\x')} ~ \min_{\y \in \A(\x)}~\twonorm{\y - \y'}\\
        & \leq \max_{\y' \in \A(\x')} ~ \min_{\y \in \A(\x)}~\twonorm{\y - \y'} \\
        & \leq \twonorm{\x - \x'}.
    \end{align*}
    In a similar manner, one can also show that 
    \begin{equation*}
        \max_{\y \in \RSet(\x)}~\min_{\y' \in \RSet(\x')}~\twonorm{\y - \y'} \leq \twonorm{\x - \x'}.
    \end{equation*}
    In conclusion, for all $\x, \x' \in \X$, we have
    \begin{equation*}
        \distH{\RSet(\x), \RSet(\x')} \leq \twonorm{\x - \x'},
    \end{equation*}
    thus completing the proof.
\end{proof}
    
\begin{lemma}
    \label{lmm:Theta_t 1-Lip in numerical sim}
    For all $t \in \{1, \ldots, T\}$ and for all $i_t \in \nodeset$, the intersection correspondence $\Theta_t^{(i_t)}$ 
    is $1$-Lipschitz under the Hausdorff distance. 
    That is, 
    \begin{equation*}
        \distH{\Theta_t^{(i_t)}(\x), \Theta_t^{(i_t)}(\x')} \leq  \twonorm{\x - \x'}, \quad \forall \; \x, \x' \in \nsp{t-1}{j}, \text{ where } i_t \in \neighbor{j}.
    \end{equation*}
\end{lemma}

\begin{proof}
    Fix $t \in \{1, \ldots, \}$ and $i_t \in \nodeset$.
    Notice that both $\nsp{t}{i_t}$ and $\RSet(\x)$ are boxes in $\X$, which implies that the intersection $\icjx{t}{i_t}{\x}$ is also a box in $\X$.
%
    Let $\twonorm{\x - \x'} = \eta$.
    From Lemma~\ref{lmm:R(x) 1-Lipschitz Numerical Simulation} we have
    \begin{equation}
        \label{eqn:dist_H(R(x), R(x'))<=epsilon}
        \distH{\RSet(\x), \RSet(\x')} \leq \eta.
    \end{equation}
    For ease of notation, define the following set
    \begin{equation*}
        \left(\RSet(\x')\right)_\eta = \{\z' \in \R^d : \min_{\y' \in \RSet(\x')}\twonorm{\z' - \y'} \leq \eta\}.
    \end{equation*}
    From the definition of the Hausdorff distance, \eqref{eqn:dist_H(R(x), R(x'))<=epsilon} implies that
$
        \RSet(\x') \subseteq \left(\RSet(\x)\right)_\eta
$    
    and
$    
        \RSet(\x) \subseteq \left(\RSet(\x')\right)_\eta.
$    
    It follows from the definition of $\Theta_t^{(i_t)}(\x)$ that
    \begin{equation*}
        \Theta_t^{(i_t)} (\x) = \RSet(\x) \cap \nsp{t}{i_t} \subseteq \left(\RSet(\x')\right)_\eta \cap \nsp{t}{i_t}.
    \end{equation*}
    Since $\RSet(\x') \cap \nsp{t}{i_t} \ne \varnothing$, Lemma~\ref{lmm:min_P = min_P_cap_Q P Q boxes} implies that for all $\z' \in \left(\RSet(\x')\right)_\eta \cap \nsp{t}{i_t}$,
    \begin{equation*}
        \min_{\y' \in \RSet(\x')} \twonorm{\z' - \y'} = \min_{\y' \in \RSet(\x') \cap \nsp{t}{i_t}} \twonorm{\z' - \y'} \leq \eta.
    \end{equation*}
    Therefore,  $\Theta_t^{(i_t)}(\x) \subseteq \left(\RSet(\x')\right)_\eta \cap \nsp{t}{i_t} \subseteq \left(\Theta_t^{(i_t)}(\x')\right)_\eta$, which further indicates that
    \begin{equation*}
        \max_{\y \in \ict{\x}}~\min_{\y' \in \ict{\x'}}~\twonorm{\y - \y'} \leq \eta.
    \end{equation*}
    Using a similar approach, one can also show that
    \begin{equation*}
        \max_{\y' \in \ict{\x'}}~\min_{\y \in \ict{\x}}~\twonorm{\y - \y'} \leq \eta.
    \end{equation*}
    Finally,  we have that
    \begin{equation*}
        \distH{\ict{\x}, \ict{\x'}} \leq \eta = \twonorm{\x - \x'}.
    \end{equation*}
\end{proof}

\end{appendices}

\end{document}